
\input phyzzx
\input epsf
\sequentialequations
\def\refmark#1{ [#1]}

\vsize =8.5truein
\hsize= 6truein
\hoffset= 0.25truein
\baselineskip=14truept
\overfullrule=0pt

\catcode`@=11
\def\@versim#1#2{\lower.7\p@\vbox{\baselineskip\z@skip\lineskip-.5\p@
    \ialign{$\m@th#1\hfil##\hfil$\crcr#2\crcr\sim\crcr}}}
\def\ga{\mathrel{\mathpalette\@versim>}} %
\def\la{\mathrel{\mathpalette\@versim<}} %
\catcode`@=12 

\def\mpl{M_{\rm Pl}}
\def\c{{\bf C}}

\def\J{{\cal J}}

\def\splus{k_+}
\def\sminus{k_-}
\def\high{\vphantom{\Biggl(}\displaystyle}
\def\rint{\int_{r_H}^\infty dr }
\def\rtint{\int dt \rint}

\def\main#1{{\vskip1truecm  \fourteenbf\noindent #1 \vskip
0.5truecm}}
\def\sub#1#2{{\vskip1truecm\noindent #1 \it  #2 \vskip
0.5truecm}}
\def\mainsub#1#2#3{\vbox{{\vskip1truecm { \fourteenbf\noindent #1 }\nobreak
\vskip 0.5truecm \nobreak\noindent#2 \it   #3 \vskip 0.5truecm}}}

\def\pr#1#2#3#4{{\it Phys. Rev. D\/}{\bf #1}, #2 (19#3#4)}
\def\np#1#2#3#4{{\it Nucl. Phys.} {\bf B#1}, #2 (19#3#4)}

\def\prl#1#2#3#4{{\it Phys. Rev. Lett.} {\bf #1}, #2 (19#3#4)}
\def\prold#1#2#3#4{{\it Phys. Rev.\/} {\bf #1}, #2 (19#3#4)}

\line{\hfil CU-TP-672}
\line{\hfil gr-qc/9503032 }
\vglue .5in

\centerline{\fourteenbf Magnetically Charged Black Holes with Hair}
\bigskip
\centerline{\twelverm Erick J. Weinberg}
\medskip
\centerline{\twelverm Department of Physics, Columbia University}
\centerline{\twelverm New York, NY 10027, USA}
\vskip .1in
\noindent\footnote{}{\twelvepoint \noindent Lectures presented at the
XIII International Symposium ``Field Theory and Mathematical Physics'',
Mt. Sorak, Korea (June-July 1994).  This work was supported in part by
the US Department of Energy.}

\bigskip
\centerline{\twelvebf Abstract}
\medskip
\centerline{\vbox{\hsize=5truein
\baselineskip=12truept
\tenrm \noindent In these lectures the properties of magnetically
charged black holes are described.  In addition to the standard
Reissner-N\"ordstrom solution, there are new types of static
black holes that arise in theories containing
electrically charged massive vector mesons.  These latter
solutions have nontrivial matter fields outside the horizon;
i.e., they are black holes with hair.
While the solutions carrying unit magnetic charge are spherically
symmetric, those with more than two units of magnetic charge
are not even axially symmetric.  These thus provide the first
example of time-independent black hole solutions that have no
rotational symmetry.  }}

\main{1. Introduction}
\nobreak

    Almost two centuries ago Laplace published his {\it Exposition du
system du
monde}\ref{P.-S.~Laplace, {\it Exposition du system du monde}, vol II,
(Paris, 1796)}.  Its final chapter contained a number of
speculations on the future of astronomy.  Among these was the
observation that ``A luminous star of the same density as the Earth,
and whose diameter was two hundred and fifty times larger than that
of the Sun, would not, because of its attraction, allow any of its
rays to reach us; it is thus possible that the largest luminous
bodies in the universe may, by this means, be invisible.''  Two
centuries later, black holes have
indeed become objects of great astrophysical interest. There is strong
evidence for their production by the collapse of dying stars and for
the existence of black holes of great size in the cores of some
galaxies.  It has also been speculated that microscopic black holes
produced shortly after the big bang may account for a portion of the
dark matter in the universe.

     However, there is a second aspect of these objects
that could not have been anticipated by Laplace.  With the advent of
special relativity, an escape velocity equal to the speed of light
indicates not merely an astronomically fascinating object but also
suggests a sharp departure from the normal structure of space-time.
This takes black holes from the realm of astrophysical phenomena into
the arena of fundamental physics.

      Closer theoretical consideration of these objects reveals
several remarkable features.  One is the occurrence of
space-time singularities.  These are first encountered  in the
simplest black hole solution, that of Schwarzschild.  In that
context, one might  plausibly guess that the singularity was an
artifact of the great symmetry of the solution and that it would
disappear with even the slightest departure from spherical symmetry.
To the contrary, it turns out that, under rather
general conditions, the classical equations describing
gravitational collapse from an initially nonsingular configuration
inevitably lead to a space-time singularity.  The precise nature of
the singularity, and the question of whether or not it is hidden by
a horizon from the view of distant observers, remain topics of active
research.  A second line of investigation follows from the
discovery\Ref\Hawking{S.W.~Hawking, {\it Commun. Math. Phys.} {\bf 43},
199 (1975).} that quantum mechanics implies that a black hole loses
mass by the emission of  thermal radiation, with the strong possibility
that this could lead to complete evaporation of the black hole.  This
process leads to a number of puzzles concerning the fate of the
information associated with matter that falls into the black hole, and
has raised the question of whether the fundamental laws of quantum
mechanics must be modified.

     I will be concerned in these lectures primarily with a third
aspect of black holes, namely the symmetry and simplicity of the
classical black hole solutions.  It is not particularly surprising
that the Schwarzschild and other known closed-form solutions have
these properties; after all, the solutions most likely to be found
analytically are those that are symmetric and algebraically simple.
What is unusual is the degree to which these properties are generic.
Consider, for example, the case of gravity coupled to
electromagnetism with no other types of matter present.  All black
hole solutions that are both time-independent and invariant under
time reversal are spherically symmetric.  If the latter condition is
relaxed, there are also solutions with nonzero angular momentum; the
direction of this angular momentum removes the isotropy, but an axial
symmetry remains.  Furthermore, these solutions are completely
specified by the associated conserved quantities --- their total
energy, angular momentum, and electromagnetic charges.  It was also
shown for a number of cases that even with the inclusion of
additional matter fields in the theory one could not produce black
holes with more structure; the failure of such attempts led to the
statement that ``black holes have no hair.''  Although these results
were all obtained for specific cases, one might have guessed that
there was in fact a more general result.  This turns out not to be
the case --- theories containing electrically charged vector mesons
admit magnetically charged black hole solutions with rather
nontrivial structure.  In these lectures I will describe these
fascinating objects and the methods by which they were found.

  I begin with two sections reviewing some standard results.
The first,  Sec.~2, contains a brief survey of some
topics in black hole physics,
with an emphasis on the properties of the classical solutions.  The
Reissner-Nordstr\"om charged black hole solution is described in this
section. I then recall, in Sec.~3, some of the properties of magnetic
monopoles in flat space-time, including in particular those of the
nonsingular topological monopoles\Ref\thooft{G.~'t~Hooft,
\np{79}{276}74; A.M.~Polyakov, {\it Pisma v. Zh. E.T.F.,} {\bf 20}, 430
(1974) [{\it JETP Lett.} {\bf 20}, 194 (1974)].} that arise in
spontaneously broken gauge theories.    Next, in Sec.~4, I describe the
effects of gravity on these monopoles and show that the equations
describing these curved space monopoles also admit a new class of black
hole solutions, carrying a single unit of magnetic charge, that have a
cloud of charged vector mesons (i.e., ``hair'') outside the horizon.  I
also show that in the context of this Higgs theory  a
Reissner-Nordstr\"om black hole carrying unit magnetic charge has a
classical instability leading to the formation of this new type of
black hole if its horizon radius is sufficiently small.  In Sec.~5
these results are extended to a more general class of theories
containing electrically charged massive vector mesons that are not
necessarily generated via the Higgs mechanism from a non-Abelian gauge
theory.  It is shown that for certain choices of parameters such
theories have finite energy monopole solutions in flat space and that,
even if these conditions are relaxed, they give rise to magnetically
charged black holes with hair.   The magnetically charged solutions
considered up to this point are all spherically symmetric and all carry
unit magnetic charge.  Black hole solutions with higher magnetic charge
are also possible.  However, essentially for reasons arising from the
anomalous angular momentum of a charge-monopole pair, these cannot
(except for the Reissner-Nordstr\"om case) cannot be spherically
symmetric.  A useful tool for investigating these nonsymmetric
solutions is the formalism of monopole scalar and vector spherical
harmonics, which is developed in Sec.~6.  In Sec.~7 this formalism is
employed to extend   the stability analysis of Sec.~4   to
Reissner-Nordstr\"om black holes in an arbitrary theory with charged
vector mesons, with no restriction on the magnetic charge. The
results of this analysis imply the existence of new black holes with
hair that, because they carry other than unit magnetic charge, are
not spherically symmetric.  While an exact closed form solution for
these objects is probably unattainable, a perturbative scheme for
constructing them can be developed under certain conditions.   This
scheme is described in Sec.~8 and is used to show that, at least for
certain choices of parameters, these new solutions are not even
axially symmetric.

\mainsub{2. Black Holes}{2.1}{The Schwarzschild Solution}
\nobreak

       The most elementary problem to be solved in Newtonian gravity is
the calculation of the gravitational field of a point mass at rest.
Let us consider the general relativistic analogue of this
problem\ref{For a fuller treatment of the material in this section, see
C.W.~Misner, K.S.~Thorne, and J.A.~Wheeler, {\it Gravitation}, (Freeman,
San Francisco, 1973); R.M.~Wald, {\it General Relativity}, (University
of Chicago, Chicago, 1984).}.   We expect the metric to be static (i.e.,
both independent of time and invariant under time reversal) and
spherically symmetric.  Any such metric can be written in the form
$$ ds^2 = - B(r) dt^2 + A(r) dr^2
   + r^2 \left(d\theta^2 + \sin^2\theta d\phi^2 \right) .
   \eqn\sphericalmetric $$
The energy-momentum tensor should vanish everywhere except at the
position of the point mass.  Hence, one should insert this form for the
metric into the source-free Einstein equations and solve, subject to
the boundary condition that the spacetime be asymptotically flat.
This leads, by well-known steps, to the
Schwarzschild\ref{K.~Schwarzschild, {\it Sitzber. Deut. Akad. Wiss.
Berlin}, Kl. Math.-Phys. Tech. 189 (1916).} metric
$$  ds^2 = - \left(1-{2MG\over r}\right) dt^2
    + \left(1-{2MG\over r}\right)^{-1} dr^2
   + r^2 \left(d\theta^2 + \sin^2\theta d\phi^2 \right)
   \eqn\Schmetric $$
where $M$ is an arbitrary constant that arises in the course of
integrating the field equations.  The meaning of $M$ is obtained by
examining the behavior of the metric at large distance.  If we recall
that for weak gravitational fields
the Newtonian gravitational potential is related to the spacetime
metric by
$$   g_{tt} = - 1 - 2 \phi_{\rm Newton}
   \eqno\eq $$
we see that $M$ is just the mass that an observer at large distances
would measure by, for example, observations of Keplerian orbits.

      This metric has singularities at $r=2M$, where
$g_{tt}=0$ and $g_{rr}=\infty$, and at $r=0$, where $g_{tt}=\infty$
and $g_{rr}=0$.  One possibility is that these are coordinate
singularities that reflect singularities in the definition of the
coordinates rather than any singularity in spacetime.  An
two-dimensional example of this is provided by the metric
$$ ds^2 = dr^2 + r^2 d\theta^2 .
      \eqno\eq  $$
Transforming from polar to Cartesian  coordinates removes the
singularity at $r=0$ and shows that the space is simply the flat
two-dimensional plane.  By contrast, for $\alpha< 1$ the metric
$$ ds^2 = dr^2 + \alpha r^2 d\theta^2
      \eqno\eq  $$
describes a cone and has a true singularity at $r=0$, as can be
verifed by a calculation of the curvature.

     The curvature scalar $R$ and the Ricci tensor $R_{\mu\nu}$
vanish everywhere in this  spacetime.  However, the
scalar $R^{\alpha\beta\mu\nu}R_{\alpha\beta\mu\nu}$ is nonzero and
equal to $48M^2G^2/r^6$,  thus showing that the metric singularity at
$r=0$ does indeed reflect a true singularity of the spacetime.

    The singularity at $r=2M$, on the other hand,
 is only a coordinate
singularity.  This can be demonstrated\ref{M.D.~Kruskal,
\prold{119}{1743}60.} by introducing Kruskal coordinates $X$ and $T$ that
are related to $r$ and $t$ by
$$ \eqalign{ X^2 -T^2 &= \left({r\over 2MG} -1\right) e^{r/2MG} \cr
       {T \over X} &= \tanh (t/4MG) .}
   \eqno\eq $$
The metric can then be written in the form
$$ ds^2 = {32M^3 G^3e^{-r/2MG} \over r} \left(-dT^2 + dX^2\right)
    + r^2 \left(d\theta^2 + \sin^2\theta d\phi^2 \right)
   \eqno\eq $$
whose only singularity is at $r=0$.

     The full range, $0\le r <\infty$, $-\infty < t<
\infty$, of the original Schwarzschild coordinates corresponds to
the region in which $T^2 -X^2 \ge 1$ and $X+T \ge 0$.  However,
since the metric is perfectly regular along the surface $X+T=0$,
there is no bar to extending the range of the Kruskal coordinates
beyond this boundary to obtain the extended Schwarzschild spacetime
shown in Fig.~1.  In this figure the coordinates have
been drawn in such a manner that light cones are $45^\circ$
lines.  The lines $X = \pm T$ divide the extended spacetime into four
quadrants.   The first, $X > |T|$, is the ``exterior'' region $r >
2M$ of the original spacetime.  The second quadrant,
$|X| < T < (X^2+ 1)^{1/2}$, corresponds to the ``interior'' region $0
< r< 2M$.
Note the causal relation between these two quadrants: a particle
can go from the first to the second quadrant by crossing the line
$X=T$ ($r=2M$), but once inside this quadrant it can never
leave.  This region is called a black hole, and its boundary the
horizon. The third quadrant, $(X^2+ 1)^{1/2} < T < -|X|$, is in a
sense the opposite of the second.  Any particle inside it must
eventually leave and can  never  return.   Finally, the fourth
quadrant, $X < -|T|$, is a second exterior region quite similar to
the first quadrant but completely inaccessible to it.

     We can see from this that the interpretation of the
Schwarzschild coordinates $r$ and $t$ is not as simple as one might
have thought when first writing down the metric \Schmetric.   Inside
the horizon, $t$ is a spatial coordinate rather than a timelike one.
Thus, although our metric is indeed independent of $t$, the
spacetime is not truly static for $r < 2M$.  Similarly, $r$ is
timelike inside the horizon, although it does continue to specify
the circumferences of the two-spheres generated by the $SO(3)$
isometries of the metric.   The ``point'' $r=0$ is in fact a time,
not a position.  Examination of Fig.~1 shows that there are in fact
many spacelike surfaces on which $r$ is never zero.   Although we
began by seeking the metric generated by a point mass at the origin, and
have indeed found a solution that at large distances is consistent with
the Newtonian potential  from such a source, our
spacetime has no static point mass inside it.

\bigskip
\centerline{\epsffile{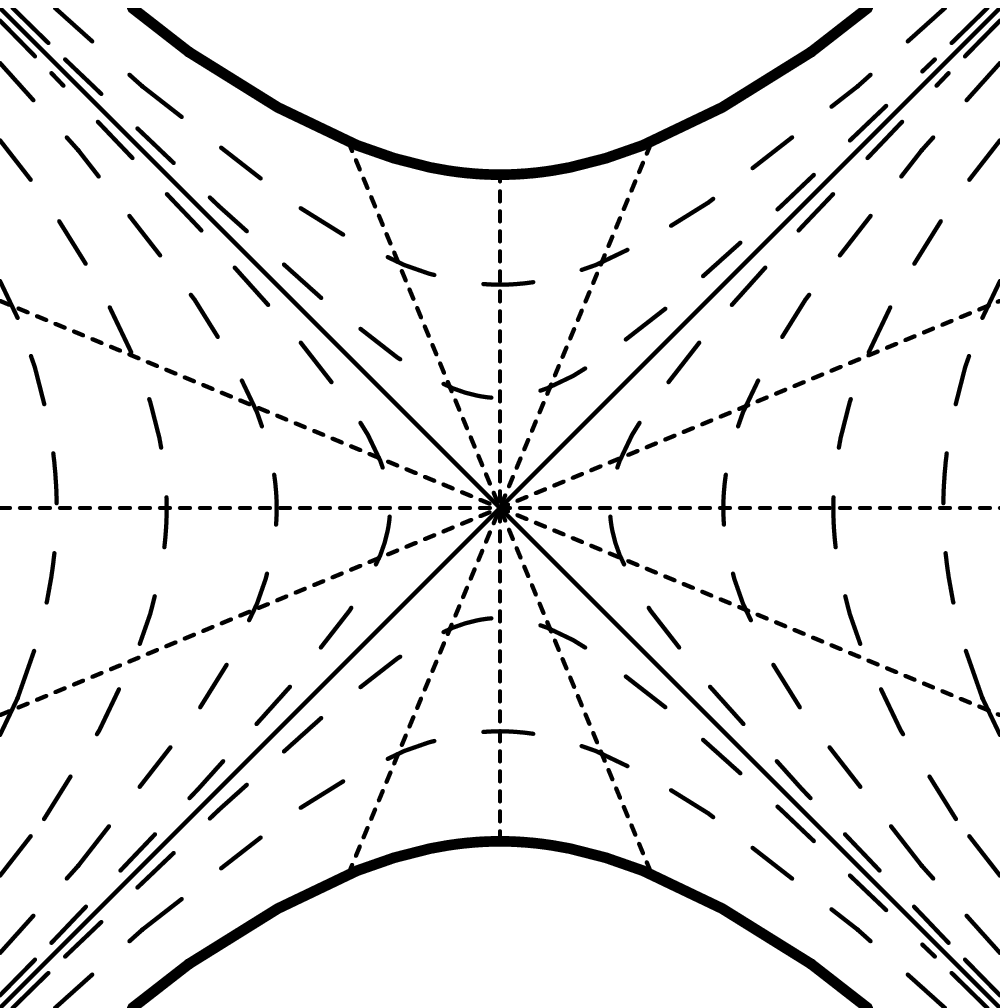} }
\bigskip
\centerline{\vbox{ \tenrm \baselineskip=12truept \hsize=4.0truein
\noindent Fig.~1.  The extended Schwarzschild spacetime.  Each point in the
plane represents a two-sphere of constant $X$ and $T$ (and thus of
constant $r$ and $t$).  The $X$ and $T$ axes, which are not
shown here, would be the usual Cartesian axes, so lightcones
correspond to $45^\circ$ lines. Lines of constant $r$
and $t$ are indicated by long and short dashes, respectively.
The singularities at $r=0$ are indicated by the heavy
solid lines, while the lighter solid lines indicate the horizon,
$r=2MG$. }}

     Because the horizon at $r=2M$ is merely a coordinate
singularity, from a purely local viewpoint one would not perceive any
distinction between points just outside the horizon and those just
inside it.  Viewed globally, of course, the causal structure of
the spacetime makes a sharp distinction between these.   To
illustrate this difference between the local and the global
viewpoints, consider two classes of
observers near the horizon.  The first falls
freely toward, and then through, the horizon.  Although they suffer
strong gravitational tidal forces as they approach the neighborhood
of the horizon, and are crushed to death in the $r=0$ singularity soon
after entering the black hole, they feel no distinct effect at the
moment of crossing the horizon.   The second are a set of static
observers who remain at fixed values of the Schwarzschild coordinate
$r$.  To keep from falling into the black hole, some external force
must be applied; one might imagine them to be hanging from long ropes
whose other ends were held by stationary
assistants far from the black hole.  For these observers the
horizon is quite special; for example, the value they measure
for the acceleration of gravity, $g$, diverges as their position
approaches the horizon.

  To the freely falling observers, the static observers at the
horizon appear to be undergoing an infinite acceleration.  From this local
point of view, the divergence in the measurements of $g$ reflects the
unusual behavior of the observer rather than any special property of
spacetime.   For the behavior of the static observers to seem anything
but bizarre, one must step back and view the spacetime as a whole.  The
static nature of the metric in the exterior regions is then apparent, and
it becomes quite reasonable to consider the measurements of an observer
with fixed Schwarzschild position.

    Finally, one must keep in mind that the full extended Schwarzschild
spacetime, although a solution of Einstein's equations, is
unlikely to occur in the real world, where black holes are expected to be
formed by gravitational collapse from nonsingular initial conditions.
After the collapse, such black holes will settle down to almost static
configurations that are well described by the Schwarzschild metric.
These asymptotic spacetimes have  exterior and interior regions that
approximate portions of the first two quadrants of Fig.~1.  The third
and fourth quadrants, however, are completely absent.

\sub{2.2}{Other Black Holes }
\nobreak

    By endowing the black hole with electric or magnetic charge, one
obtains the spherically symmetric Reissner-Nordstr\"om
solution\ref{H.~Reissner, {\it Ann. Physik} {\bf 50}, 106 (1916);
G.~Nordstr\"om, {\it Proc. Kon. Ned. Acad. Wet.} {\bf 20}, 1238 (1918).}.
This  has radial electric and magnetic fields
$$ \eqalign{ E_r &= {Q_E\over r^2} \cr
      B_r &= {Q_M \over r^2} }
    \eqno\eq $$
and a metric
$$ ds^2 = - B_{RN}(r) dt^2 + B^{-1}_{RN}(r) dr^2
   + r^2 \left(d\theta^2 + \sin^2\theta d\phi^2 \right)
   \eqno\eq $$
where
$$ B_{RN}(r) = 1 -{2MG \over r} + {4\pi G \sqrt{Q_E^2 + Q_M^2} \over r^2}
\,.
    \eqno\eq $$

    The nature of this spacetime depends on the value of the mass
$M$.  If $M$ is greater than the extremal mass
$$ M_{\rm ext} = \sqrt{4\pi (Q_E^2 + Q_M^2)}\, \mpl
    \eqno\eq $$
(where $\mpl = G^{-1/2}$ is the Planck mass),
the metric has coordinate singularities at the
zeroes of $B(r)$, which occur at
$$ r_\pm = MG \left[ 1 \pm
   \sqrt{1 - \left({M_{\rm ext} \over M}\right)^2 }\,\,\right] \, .
    \eqno\eq $$
In addition, there is a true physical singularity at $r=0$.   As with
the Schwarzschild solution, this spacetime can be extended beyond the
region covered by the original coordinates $r$ and $t$.  The causal
structure of this extended spacetime is more complex than in the
Schwarzschild case.  However, as in that case, there is a horizon at
$r=r_+$ with the property that objects that cross inside it
can never return to the exterior region $r_+ < r < \infty$.

     If $M=M_{\rm ext}$, so that the two zeroes of $B_{RN}(r)$ coincide,
one obtains the extremal Reissner-Nordstr\"om black hole.   There is
still a horizon, but the double zero in $g_{rr}^{-1}$ implies that it
lies at an infinite proper distance from any point in the exterior
region $r> r_H$.   One also has the curious fact that there is no
long-range static force between two extremal Reissner-Nordstr\"om black
holes with charges of the same sign, because their mutual gravitational
attraction is exactly cancelled by their electrostatic and magnetostatic
repulsion.  (In fact, static multi-black hole solutions for the extremal
case are known\ref{S.D.~Majmudar, \prold{72}{390}47; A.~Papapetrou, {\it
Proc. Roy. Irish Acad.} {\bf 51}, 191 (1947). }.)

     The final possibility is $M< M_{\rm ext}$.  In this case one
has a ``naked singularity'' in which there is no horizon shielding
the asymptotic region from the effects of the physical singularity at
$r=0$. (The Schwarzschild solution with negative $M$ also has a naked
singularity.)   For the remainder of these lectures I will exclude
naked singularities from the discussion.  Indeed, it has been
conjectured that there is a ``cosmic censorship'' that prevents the
evolution of a naked singularity from  generic nonsingular initial
conditions.

     There is a simple physical explanation for the existence of a
minimum horizon radius about a charged black hole.  Roughly speaking,
one expects a horizon to form whenever a region of radius $r$ contains a
total energy much greater than $r/G$.  If a charged black
hole were to have a horizon at $r_H \ll QG^{1/2}$, the energy in the
Coulomb field just outside the horizon (say, in the region  $r_H <
r < 2r_H$) would be much greater than $r_H/G$, thus implying the
existence of a horizon at a radius much greater than the hypothesized
$r_H$.

     The Schwarzschild and Reissner-Nordstr\"om solutions are both
static.  It is also possible to have a metric that is stationary
(i.e., time-independent) but, because of the presence of nonzero
time-space components, not invariant under time reversal.  An example
of this is the axisymmetric Kerr metric\ref{R.P.~Kerr,
\prl{11}{237}63.},  which describes a black hole with a mass $M$ and an
angular momentum $J$ along the axis of symmetry; it can be
generalized\ref{E.T.~Newman et al, {\it J. Math. Phys.} {\bf 6},
918 (1965).} to include the possibility of nonzero electric and magnetic
charges.  As with the previous black hole solutions, there is a region,
bounded by a horizon, from which one cannot escape to the asymptotically
flat external region.  However, in constrast with these solutions, the
surface on which $g_{tt}$ vanishes lies outside the horizon, meeting it
only on the axis of symmetry. The region lying between these two surfaces
is known as the ergosphere (the significance of this name will be
explained below).  Within the ergosphere a particle cannot be stationary,
but must have an angular velocity greater than an ($r$-dependent) minimum.

      One might well expect there to be additional types of black
holes.  For example, one might seek solutions that corresponded at
large distances to the Newtonian gravitational field due to a mass
distribution with higher multipole moments.  Remarkably, if we
restrict ourselves to gravity plus electromagnetism, there are no more
black holes; the only time-independent solutions  are those described
above.  These can be characterized completely by their mass, angular
momentum, and electromagnetic charges.  They are all at least axially
symmetric, and the static solutions are all spherically symmetric.

      The next step, of course, is to see if more structure can be
obtained by including other types of matter fields in the
theory.   This possibility has been  ruled out for a number of
cases,  giving rise to the statement that ``a black
hole has no hair''.   However, it should be realized that there is no
general no-hair theorem but only a collection of theorems
applicable to very specific cases.  As an example of these,
consider the case\ref{J.D.~Bekenstein, \pr{5}{1239}72.} of a
massive scalar field $\phi$ with
the matter portion of the action taking the form
$$ S_{\rm matter} = \int d^4x \sqrt{g}
  \left[{1\over 2}\partial_\mu\phi \partial^\mu\phi - V(\phi)
   \right] \, .
    \eqno\eq $$
For the sake of simplicity, let us assume spherical symmetry, which
implies that the metric can be written in the form \sphericalmetric.
For a static solution, the field equations take the form
$$  {1\over r^2\sqrt{AB}}{d\over dr} \left[ {r^2\sqrt{AB} \over A}
\phi'\right]
     = {dV\over d\phi} \, .
    \eqno\eq $$
where a prime denotes a derivative with respect to $r$.
Let us now multiply both sides of this equation by $\phi -\phi_0$, where
$\phi_0$ is the value which minimizes $V(\phi)$.  After some
rearranging, we obtain
$$ (\phi-\phi_0) {d\over dr}\left[ {r^2\sqrt{AB} \over A} \phi'\right]
   = (\phi-\phi_0) r^2\sqrt{AB}\, {dV\over d\phi} \, .
    \eqno\eq $$
Integrating this over $r$ from the horizon to infinity leads to
$$  \int^\infty_{r_H} dr \, {d\over dr} \left[ {r^2\sqrt{AB} \over A}
   (\phi-\phi_0) \phi' \right]
   =  \int^\infty_{r_H} dr\, r^2\sqrt{AB} \left[ {(\phi')^2 \over A}
     + (\phi-\phi_0){dV\over d\phi} \right] \, .
    \eqno \eq $$

    The left-hand side of this equation can be immediately integrated
to give two surface terms.  The one from $r=\infty$ vanishes because
$\phi$ tends asymptotically to its vacuum value, $\phi_0$, with the
deviation from $\phi_0$ falling exponentially fast because we are
dealing with a massive field.  The surface term from $r=r_H$ is also
zero, since $g^{rr}= A^{-1}(r)$ vanishes on the horizon.  On the
right-hand side, the contribution from the first term in brackets is
clearly positive, since $A(r)>0$ for all $r>r_H$.  The contribution due
to the second term is also positive {\it if} we assume that $\phi_0$ is
the only minimum of $V$.  When this is the case, we have two positive
terms whose sum must vanish.  This implies that the two vanish
separately, and allows only the trivial solution in which $\phi$ is
identically equal to its vacuum value everywhere outside the horizon.
Note that this argument fails if $V(\phi)$ has minima other than
$\phi_0$, even if these are only local minima.

\sub{2.3}{Energy Extraction and Black Hole Thermodynamics}
\nobreak

      It is clear that the mass of a black hole can increase by absorption
of infalling matter.  Given the fact that particles cannot escape from
inside the horizon, it might seem equally clear that the mass can never
decrease.   In fact, this is not so.  One example of such a process
is associated with the Kerr black hole.   Recall that there is a region
just outside the horizon, known as the ergosphere, in which $g_{tt}$ is
positive (i.e., of the nonstandard sign).   Using this fact, one can
show that it is possible for a particle in the ergosphere to have a
negative total energy; very roughly speaking, one might say that it has
a negative gravitational potential energy that is larger in magnitude
than its rest energy.  Several mechanisms for extracting energy from the
black hole then suggest themselves.  For example, one might send a
positive-energy object into the ergosphere, have it produce some
negative-energy particles that are then sent inside the horizon, and
then have the object emerge from the ergosphere with more energy than it
started with.   The net effect of this would be to reduce the mass of
the black hole.

    This process cannot be repeated indefinitely.  Particles inside the
ergosphere cannot be static, but must instead have a minimum angular
velocity whose sign is such that any negative-energy particles
falling inside the horizon decrease the total angular momentum of the
hole.  This in turn reduces the size of the ergosphere.  Eventually, the
hole loses all of its angular momentum, thus becoming a Schwarzschild
black hole, and the possibility of energy extraction ceases.

     Energy can also be extracted from a Reissner-Nordstr\"om black hole.
To see this, consider a particle near the horizon that carries a charge of
the same type (i.e., electric or magnetic) but opposite sign as the black
hole.  If the charge to mass ratio of the particle is large enough, its
negative potential energy can be of
sufficient magnitude to cancel the rest mass and give the particle a
negative total energy.   As with the Kerr example, one can use such
negative-energy particles to reduce the mass of the black hole.  One
simple mechanism is to produce a pair of particles with opposite
sign near the horizon, letting the one with negative energy fall through
the horizon while the one with positive energy goes off to spatial
infinity.  Again, this process is self-limiting, in that the
negative-energy particles reduce the charge of the black hole, with energy
extraction no longer possible once the hole is completely neutralized.

     Further analysis leads to the formulation of several laws of black
hole dynamics that govern these and other processes:
\item{0.}  One can define a quantity $\kappa$, known as the surface
gravity, that is constant over the horizon of a stationary black hole.
\item{1.}  The change in black hole mass can be written as a sum of a
term proportional to the change in the area $A$ of the horizon and
``work terms''.  In the case of a Kerr black hole, for example, one has
$$ dM = {\kappa \over 8\pi} dA + \Omega_H dJ \, ,
   \eqno \eq $$
where $\Omega_H$ may be interpreted as the angular velocity of the
horizon.
\item{2.}  The area of the black hole horizon can never decrease; i.e.,
$$ dA \ge 0 \, .
    \eqno \eq $$
\item{3.}  One cannot reach the limit $\kappa=0$ by any physical process.

    These are remarkably similar in form to the corresponding laws of
thermodynamics.   To convert one set to the other, we need only
replace the work term in the first law by $PdV$ and make
the substitutions  $\kappa \rightarrow T/a$, $M \rightarrow E$, and $A
\rightarrow 8\pi a S$, where  $T$, $E$, and $S$ are the temperature,
energy, and entropy, respectively, and $a$ is an arbitrary
constant.  Hawking's discovery\refmark{\Hawking} that quantum
mechanically black holes  radiate with a temperature $T_H = \hbar\kappa
/2\pi$ fixes the constant $a$, and indicates a deeper meaning to this
analogy.

     There are a several ways to determine the Hawking temperature
of a black hole.  A particularly convenient method\ref{G.W.~Gibbons and
S.W.~Hawking, \pr{15}{2752}77.}
is based on the formulation of thermodyanmics in terms of a
Euclidean path integral over configurations with
periodicity $\hbar/T$.  To apply this to a spherically symmetric
spacetime with a metric of the form \sphericalmetric, we first make the
replacement $t \rightarrow i\tau$, thus obtaining the Euclidean metric
$$ ds_E^2 = B(r)d\tau^2 + A(r) dr^2 + r^2 d\theta^2
     + r^2\sin^2\theta  d\phi^2 \, .
      \eqno \eq $$
We want $\tau$ to be a periodic variable with period $\hbar/T$; this
suggests that we define an angle
$$  \alpha ={ 2\pi T \over \hbar} \tau
    \eqno \eq$$
that is understood to have period $2\pi$.
If we also change variables from $r$ to a quantity
$$  R(r) = {\hbar \over 2\pi T} \sqrt{B(r)} \,\, ,
    \eqno\eq $$
the Euclidean metric takes the form
$$  ds_E^2 = {4 AB \over (B')^2} \left({4\pi T\over \hbar}\right)^2 dR^2
      +R^2 d\alpha^2 + r^2\sin^2\theta  d\phi^2 \, .
      \eqn\newEuc $$
The horizon of the Lorentzian signature spacetime lies at the zero of
$B(r)$.  In the Euclidean metric \newEuc\ this corresponds to the zero
of $R(r)$, which may be viewed as the origin of a two-dimensional subspace
of constant $\theta$ and $\phi$ that is
spanned by polar coordinates $R$ and $\alpha$.  The
temperature is fixed by requiring that the singularity at this point be
merely a coordinate singularity rather than a true conical singularity.
This is achieved by arranging that the coefficient of $dR^2$ be equal
to unity at $r=r_H$, which gives
$$  T = {\hbar \over 4\pi} \left({B'\over \sqrt{AB}} \right)_{r=r_H} \, .
      \eqn\Htemp $$

      For the Schwarzschild metric, we obtain a temperature
$$   T_H = {\hbar \over 8\pi MG}
     \eqno\eq $$
that increases monotonically as the mass decreases.   For the
Reissner-Nordstr\"om black hole the temperature is
$$    T_H =  {\hbar\over 2\pi G}{\sqrt{M^2 -M^2_{\rm ext}} \over
       \left[ M + \sqrt{M^2 -M^2_{\rm ext}}\, \right]^2 } \, .
    \eqno\eq $$
For large $M$ this also increases with decreasing mass.  However, as
the mass approaches its extremal value the temperature turns over and
begins to decrease, finally tending to zero as
$M\rightarrow M_{\rm ext}$.

     The Hawking radiation removes mass from the black hole
at a rate proportional to the area of the horizon times the fourth power
of the temperature.  In the Schwarzschild case, this is proportional to
$1/M^2$, suggesting that the black hole  evaporates
completely in a time
$$  t_{\rm evap} \sim \left({M\over M_{\rm Sun}}\right)^3 (10^{71} \,\rm
seconds )\, .
   \eqno \eq $$
For the Reissner-Nordstr\"om case, on the other hand, the evaporation
turns off as the black hole approaches the extremal limit, assuming that
the hole has not in the meantime
been discharged by some process.

     It must be kept in mind, of course, that the semiclassical
approximation used to derive the Hawking process breaks down as the
black hole approaches Planck size.   Whether complete evaporation actually
occurs and whether this leads to information loss and a breakdown of the
unitary evolution of quantum mechanics are important issues that lie
beyond the scope of these lectures.

\main{3.  Magnetic Monopoles in Flat Spacetime}
\nobreak

     Let us now recall some facts about magnetic monopoles in flat
space-time.   A monopole with magnetic charge $Q$ gives rise to a
radial magnetic field
$$  {\bf B} = {\bf \hat r} {Q \over r^2}
   \eqno \eq $$
that can be derived from the Dirac vector potential with components
$$  \eqalign{ A_r^{\rm Dirac} &= 0 \cr
     A_\theta^{\rm Dirac} &=0 \cr
     A_\phi^{\rm Dirac} &= Q(1-\cos\theta) \, .}
    \eqn\Diracpotential $$
This potential has a singularity (``the Dirac string'') along the
negative z-axis.  Electromagnetic gauge-transformations can change the
orientation of this singularity, but cannot remove it.  By requiring
that the string be unobservable (for example, by Aharonov-Bohm
experiments) one obtains the Dirac quantization
condition\ref{P.A.M.~Dirac, {\it Proc. Roy. Soc. London\/} {\bf A133},
60 (1931). }  $$  Q ={q\over e}
   \eqno\eq $$
where $e$ is the smallest electric charge in the theory and $q$ is
either an integer or an integer plus one half.

     Even when the Dirac quantization condition is satisfied, a
singularity at the origin remains; this is, of course, no more
remarkable than the singularity of a point electric charge.  However,
it was shown by 't~Hooft and Polyakov\refmark{\thooft} that
certain spontaneously broken gauge theories have completely nonsingular
classical solutions that carry magnetic charge.  The simplest case is an
$SU(2)$ gauge theory that is spontaneously broken to $U(1)$ when a triplet
Higgs field $\phi$ acquires a nonzero vacuum expectation value of
magnitude $v$; I will use the language of electromagnetism to describe
this unbroken $U(1)$.  This theory is governed by the Lagrangian
$$ {\cal L} = -{1 \over 4} F_{\mu\nu}^a F^{a\mu\nu}
    +{1\over 2} (D_\mu \phi)^a (D^\mu \phi)^a
     - {\lambda\over 2}(\phi^a\phi^a -v^2)^2
    \eqn\Higgslag $$
where the field strength
$$  F_{\mu\nu}^a = \partial_\mu A^a_\nu -\partial_\nu A^a_\mu
        - e \epsilon_{abc} A_\mu^b A^c_\nu
    \eqno\eq $$
and the covariant derivative
$$ D_i \phi^a = \partial_i \phi^a - e \epsilon_{abc} A_i^b \phi^c \, .
    \eqno\eq $$

   In the vacuum, $\phi$ may be chosen to have a
constant direction in internal space; to be definite, let $\phi^a =
v\delta^{a3}$.  The particle content of the theory then includes a
massless photon (corresponding to the 3-component of the gauge field),
two charged vector particles with mass $ev$ and charges $\pm e$
(corresponding to linear combinations of the 1- and 2-components of the
gauge field), and a massive neutral scalar.

     The monopole solution is obtained by allowing the orientation of
$\phi$ in internal space to vary from point to point.  Specifically, let
us suppose that the direction of $\phi$ in internal space is aligned with
the direction from the origin in physical space, so that $\phi^a
\rightarrow v{\hat r}^a$ as $r \rightarrow \infty$.  Configurations
with this asymptotic behavior are topologically distinct from the
vacuum; i.e., they cannot be smoothly deformed into a configuration
where $\phi$ has  a uniform direction.  Minimizing the energy among
configurations of this topological class gives a new static
solution.  To see that this solution is a magnetic monopole, we must
examine the long-range behavior of the gauge fields.  First, note that
the total energy is finite only if   $ D_i \phi$ vanishes
asymptotically.  Given the asymptotic form of $\phi$, this means that,
up to a $U(1)$ gauge transformation, the spatial components of the
asymptotic gauge field must be
$$ A_i^a = \epsilon_{iak}{\hat r}_k {1\over er}  \,  .
   \eqno \eq $$
{}From this we find that at large distance the
magnetic components of the  field strength are
$$  F_{ij}^a = \epsilon_{ijk}{\hat r}_k {\hat r}^a {1\over er^2}\, ,
    \eqno \eq $$
which is just the Coulomb magnetic field of a monopole of magnetic
charge $1/e$.

     This charge is twice the minimum value allowed by the Dirac
quantization condition.  Indeed, analysis of the possible topologies
of the Higgs field at spatial infinity shows that nonsingular
configurations can only have integral magnetic
charges\rlap.\footnote{1}{One way of understanding this is to realize
that one could easily introduce an $SU(2)$ doublet of scalar fields
into the theory, with the interactions arranged so that the unbroken
gauge group remained $U(1)$ and so that the `t~Hooft-Polyakov monopole
remained a classical solution. After symmetry breaking, this
components of the doublet would have electric charges $\pm e/2$, and so
the Dirac condition would become $Q= n/e$, with $n$ an integer.}
Although this would seem to allow nonsingular magnetic monopoles with
multiple charge, further analysis of the the field equations shows
that such solutions are absent, except in the Prasad-Sommerfield limit,
$\lambda \rightarrow 0$\ref{M.K.~Prasad and C.M.~Sommerfield,
\prl{35}{760}75.}.

    The monopole solution can be studied in further detail by
introducing the spherically symmetric ansatz
$$ \eqalign{  \phi^a &= {\hat r}^a v h(r)  \cr
     A_i^a &= \epsilon_{iak} {\hat r}_k
     \left[ {1-u(r) \over er}  \right] \, . }
     \eqn\monopoleansatz $$
The requirement that the fields be nonsingular at the origin gives
the boundary conditions
$$  h(0)=0, \qquad\qquad u(0)=1,
    \eqno\eq $$
while the asymptotic behavior described above implies that
$$  h(\infty)=1, \qquad\qquad u(\infty)=0 .
    \eqno\eq $$

    It is instructive to apply an  $SU(2)$ gauge transformation that
makes the orientation of $\phi$ spatially uniform.  This reintroduces
the Dirac string, but has the advantage of making clearer the
correspondence between the physical fields and the radial functions in
the ansatz of Eq.~\monopoleansatz.   The configuration that results takes
the form
$$   \eqalign{ \phi^a &= \delta^{a3} v h(r) \cr
     A_i^{\rm EM} &\equiv A_i^3 = A_i^{\rm Dirac} \cr
     W_i &\equiv {1\over \sqrt{2}} (A_i^1 + i A_i^2) = f_i(\theta,\phi)
{u(r)\over er} }
   \eqn\unitaryform $$
where the $f_i(\theta,\phi)$ are functions whose detailed form will not
concern us at present.   We see that $u(r)$ determines the magnitude of
the charged vector field $W_i$, while, as could already be seen from
Eq.~\monopoleansatz, $h(r)$ gives the magnitude of the Higgs field.

    Inserting the ansatz of Eq.~\monopoleansatz\ into the fields
equations leads to a pair of coupled differential equations. In general,
these can only be solved numerically.  Even without doing so, we can
obtain a rough picture of the solution.   Near the origin there is a
monopole core of radius $R_{\rm mon}$ within which $u$ is nonzero and
$h$ is close to unity.  Outside this core the two radial
functions, each of which corresponds to a massive field, tend
exponentially fast to their asymptotic values, leaving only the
Coulomb magnetic field.  If we approximate the energy density within
the core as a constant, $\rho_0$, then the total energy of the
solution is
$$  M_{\rm mon} \approx {4\pi\over 3}\rho_0 R_{\rm mon}^3
  + {2\pi\over 3}{Q_M^2\over R_{\rm mon} }
    \eqno \eq $$
where the first term is the core contribution and the second is the
energy in the Coulomb field outside the core.  Adjusting the core
radius to minimize the energy, we obtain
$$    R_{\rm mon} \sim {\sqrt{Q_M} \over \rho_0^{1/4} }
   \eqno \eq  $$
which gives
$$   M_{\rm mon} \sim Q_M^{3/2} \rho_0^{1/4} \, .
    \eqno \eq $$
If the vector and scalar masses are comparable (i.e, if $\lambda \sim
e^2$), then a reasonable estimate of the core energy density is
$\rho_0 \sim \lambda v^4 \sim e^2 v^4$.  Substituting this into the
above equations and using the fact that $Q_M = 1/e$, we obtain
$$   R_{\rm mon} \sim {1\over ev}
    \eqn\monoradius $$
and
$$  M_{\rm mon} \sim {v\over e} \, .
    \eqno \eq $$
More detailed calculations show that these estimates are in fact valid
even if $\lambda$ is not comparable to $e^2$.  In particular,
$$   M_{\rm mon} = {4\pi v\over e}  f\left( {\lambda \over e^2}\right)
    \eqno \eq $$
where $f(0)=1$ and $f(\infty) = 1.787$\ref{T.~Kirkman and C.K.~Zachos,
\pr{24}{999}81.}.

\mainsub{4.  Magnetic Monopoles in Curved Spacetime}{4.1}{Nonsingular
Magnetic Monopoles}
\nobreak

     Let us now examine the effects of gravity on the monopole
\hbox{solution\REFS\van{P.~van Nieuwenhuizen, D.~Wilkinson, and M.J.~Perry,
\pr{13}{778}76.}\REFSCON\lnw{K.~Lee, V.P.~Nair
 and E.J.~Weinberg, \pr{45}{2751}92.}
\REFSCON\maison{P.~Breitenlohner, P.~Forg\'acs, and D.~Maison,
\np{383}{357}92;  Max-Planck-Institut preprint MPI-PhT/94-87 (1994).}
 \REFSCON\ortiz{M.E.~Ortiz,
\pr{45}{2586}92.}\refsend.}\ \    In particular, consider what
happens if the
monopole mass is increased.   The Schwarzschild radius of the monopole is
$$   R_{\rm S} = 2M_{\rm mon} G \sim {v\over e M_{\rm Pl}^2} \, .
    \eqn\monoSch $$
If the radius of the monopole core were to be much less than this, we
would expect the
monopole to become a black hole.  Comparison of Eqs.~\monoradius\ and
\monoSch\ suggests that this would happen if $v$ exceeded a critical
value of order $M_{\rm Pl}$.   For such values the monopole mass
exceeds the Planck mass by a factor of roughly $1/e$.  However,
because the energy density within the monopole is of order $e^2 M_{\rm
Pl}^4$, we should be able to safely neglect quantum gravity effects
as long as $e^2$ is sufficiently small.

    To see how this works, let us examine the curved space solutions
more closely.  If we assume that the solution is static and
spherically symmetric, then the metric can be written in the form of
Eq.~\sphericalmetric, while the matter fields can be described by the
ansatz of Eq.~\monopoleansatz, suitably modified for a curved space.  The
action then reduces to
$$  S_{\rm matter} = -4\pi \int dt\, dr\, r^2 \sqrt{AB}
   \left[{K \over A}  +  U \right ]
   \eqn\matteraction $$
where
$$   K = {(u')^2 \over e^2 r^2} + {1\over 2} v^2 (h')^2
    \eqno\eq $$
and
$$  U = {(1-u^2)^2 \over 2e^2 r^4 } + {u^2h^2v^2 \over r^2}
       + {\lambda \over 2} v^4 (1-h^2)^2 \, .
    \eqno \eq $$
Thus, the theory has been effectively reduced to a theory of two
scalar fields in one space and one time dimension, but with a slightly
unconventional gradient term and with a scalar field potential
$U(u,h)$ that is position dependent.   At large $r$ the potential is
minimized by setting $u=0$ and $h=1$, while at small $r$ its minimum
is at $u=1$, $h=0$.  More precisely, the former configuration is at
least a local minimum for $1/(ev) \le r < \infty$, while the latter is
a local minimum for $0\le r \le 1/(\sqrt{\lambda} v) $.  If $\lambda <
e^2$ these two intervals overlap; the large distance minimum becomes
the global minimum at $r=1/(\lambda e^2)^{-1/4} v$.  If $\lambda >
e^2$, the two intervals are disjoint; in the region between them there
is an $r$-dependent minimum that interpolates between the
short-disance and large-distance minima.

    The field equations that follow from variation of the matter
action are
$$ \eqalign{ {1\over \sqrt{AB}}{d\over dr} \left[ {\sqrt{AB} u'\over A} \right]
    &= {e^2r^2 \over 2} {\partial U\over \partial u} \cr
    &= {u(u^2-1) \over r^2} + e^2u h^2v^2 }
    \eqn\ueq $$
and
$$ \eqalign{ {1\over r^2\sqrt{AB}}{d\over dr}
      \left[ {r^2\sqrt{AB} h'\over A} \right]
    &= { 1\over v^2} {\partial U\over \partial h} \cr
    &= {2u^2v^2h \over r^2} + \lambda v^4 h(h^2-1)\, . }
    \eqn\heq $$
These must be supplemented by Einstein's equations.   If we define a
quantity ${\cal M}(r)$ by
$$ A^{-1}(r) = 1 - {2 {\cal M}(r) G \over r} \, ,
    \eqn\calmdef$$
these may be reduced to the two equations
$$ { \cal M}' = 4\pi r^2 \left({K(u,h) \over A} + U(u,h) \right)
    \eqn\calmeq $$
and
$$  {(AB)' \over AB} = 16\pi G r K(u,h) \, .
    \eqn\ABeq $$
The last of these equations can be used to eliminate $AB$ from the
first three.  This leaves two second-order and one first-order
equation for the functions $h$, $u$, and $\cal M$.  These require five
boundary conditions.  Four are provided by the conditions
obtained above on the values of $h$ and $u$ at the origin and at
spatial infinity.  The fifth follows from noting that the metric
is nonsingular at $r=0$ only if ${\cal M}(0)=0$.  Note that ${\cal
M}(\infty)$ is unconstrained; it gives the total mass of the monopole.

     This set of differential equations can be solved numerically.  It
is interesting to compare the behavior of the solutions as the
quantity $v/M_{\rm Pl}$ is increased while the other parameters are
held fixed\refmark{\lnw-\ortiz}; this corresponds to increasing the effects of
gravity. One finds that the monopole core (as measured by the behavior of
$h(r)$ and $u(r)$) is pulled in to smaller values of $r$, until a
limiting case is reached for a critical value of $v$ of order $M_{\rm
Pl}$. For this limiting case, the variation of $h(r)$ and $u(r)$ is
confined entirely to the range $0 < r < r_{\rm ext}$, where $ r_{\rm
ext} = (\sqrt{4\pi}/e) \mpl^{-1}$ is the horizon radius of the
extremal Reissner-Nordstr\"om black
hole with unit magnetic charge; for all $r > r_{\rm ext}$, $h(r)=1$
and $u(r)=0$.  Similarly, the metric is precisely that of the extremal
Reissner-Nordstr\"om solution for $r > r_{\rm ext}$, but is smooth and
nonsingular for $0 < r < r_{\rm ext}$.  For values of $v/M_{\rm Pl}$
greater than this critical value, we find no nonsingular monopole
solutions.

    This description suggests a monopole core being pulled inward by
an increasingly strong gravitational force.  A somewhat different
picture is obtained if, instead of plotting the solutions as functions
of the coordinate $r$, one plots them as functions of the invariant
proper distance from the origin, $\ell(r) = \int_0^r dr g_{rr}^{1/2}
$.  Plotted this way, the radial functions $h$ and $u$ show very
little variation as the strength of gravity is increased.  This
remains true even when the limiting case is reached.  This is possible
because in the extremal case the double zero of $g_{rr}^{-1}$ implies
that the position of the zero lies an infinite proper distance from
the origin.

\sub{4.2}{Magnetically Charged Black Holes}
\nobreak

    Let us now turn to the magnetically charged black holes of this
theory.  First of all, we can trivially extend the Reissner-Nordstr\"om
black hole to this theory.  From examining the field equations, it is
clear that a solution can be obtained by taking the metric to be exactly
the same as in the electromagnetic case, and setting $u(r)=0$ and
$h(r)=1$ everywhere.   As before, to avoid a naked singularity we must
require that the black hole mass $M \ge M_{\rm ext}$.

    In addition, we can obtain an entirely new type of magnetic black
hole\refmark{\lnw-\maison}\ by making only a slight variation on the
procedure used to obtain the nonsingular curved space monopoles.  To be
specific, suppose that we allowed ${\cal M}(0) $ to be nonzero and
positive.  This would give a singularity at $r=0$ that we would expect to
be hidden behind a horizon at some $r_H \sim {\cal M}(0)\,G $.  If ${\cal
M}(0) $ were small enough that ${\cal M}(0) G \ll 1/(ev)$, we might thus
obtain an object that could be viewed as a tiny Schwarzschild black hole
in the center of an 't~Hooft-Polyakov monopole.  This would be a black
hole surrounded by nontrivial massive charged vector and Higgs fields; in
other words, a black hole with hair.

    To understand heuristically how the no-hair theorems are evaded,
recall that the proof given above depended on the fact that the scalar
potential had only a single minimum.  In the context of our
spherically symmetric ansatz, the Higgs theory reduces to one
involving two scalar fields governed by a potential $U(u,h)$ that
has different minima at small and large $r$.  This suggests that a
black hole with hair might exist provided that the horizon lies in the
region where $U$ is minimized by the short-distance minimum.  Indeed,
one can show that a necessary condition for such a solution is that
$r_H$ be less than the maximum of $1/(ev)$ and $1/(\sqrt{\lambda}v)$.

    We can proceed further by returning to Eqs.~\ueq-\ABeq.  Since we
have a curvature singularity at $r=0$, there is no longer any reason
to fix the values of $h(0)$ and $u(0)$.  These boundary
conditions at the origin are replaced by new boundary
conditions at the horizon.  These arise because the coefficients of
$u''$ and $h''$ in Eqs.~\ueq\ and \heq\ vanish at the zeroes of $1/A$.
Requiring that the matter fields be nonsingular at $r_H$ then gives
two constraints (one for each equation) involving the values of the
fields and their derivatives at the horizon.   If there were more than one
horizon, there would be two such conditions at each horizon.  Together
with the conditions at spatial infinity this would give at least six
boundary conditions, which would be too many for our set of
differential equations.  Hence, apart from exceptional cases (e.g.,
the Reissner-Nordstr\"om solution) we should not expect to find
solutions with multiple horizons.
(The existence of boundary conditions at the horizon,
even though there is no local physical singularity there,
reflects the global nature of the requirement we have imposed on our
solutions, namely, that they be static.)

    It is straightforward now to integrate the differential equations
numerically.  However, a somewhat different approach may be more
instructive.  By integrating Eq.~\calmeq\ we obtain
$$  {\cal M}(r) = {\cal M}(0) + 4\pi \int_0^r dr r^2 \rho(r)
    \eqno\eq $$
where $\rho = K/A + U$.  We would expect $\rho$ to be roughly constant
out to $r \sim R_{\rm mon} \sim 1/(ev) $, and then to fall as $1/r^4$.
The behavior that this implies for ${\cal M}(r)$ is indicated \break
{\epsfxsize=5.5truein
\epsfysize=2.5truein
\centerline{\epsffile{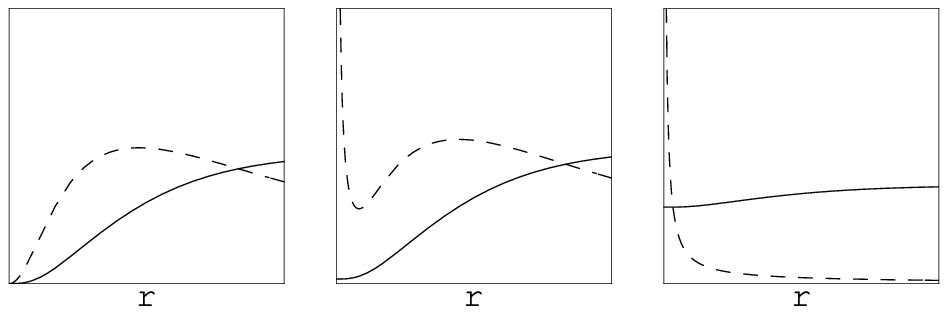}} }
\centerline{\vbox{\hsize=5.5 truein \tenrm \baselineskip=12truept
\noindent Fig.~2.  Behavior of ${\cal M}(r)$ (solid line) and ${\cal M}(r)/r$
(dashed line) for the three cases ${\cal M}(0)=0$, $0<{\cal M}(0) \ll
M_{\rm mon}$, and ${\cal M}(0) \gg M_{\rm mon}$. }}

\noindent by the solid
lines  in Fig.~2  for the three
cases ${\cal M}(0) =0$, $0< {\cal M}(0) \ll M_{\rm mon}$, and ${\cal
M}(0) \gg  M_{\rm mon}$.  In all three cases the
graph begins to level off at $r  \sim 1/(ev) $.
The dashed lines in these plots show the corresponding behavior for ${\cal
M}(r)/r$;  Eq.~\calmdef\ shows that there will be a horizon
whenever this quantity is equal to $1/(2G)$.  For the first case
(${\cal M}(0)=0$), the plot of ${\cal M}/r$ has a single peak.  For
small values of $v$ the height of this peak is less than $1/(2G)$, and
there is no horizon. As $v$ is increased, this peak  eventually
becomes high enough that a horizon forms; this is just the critical
case described previously, for which we found that the total mass of
the system was precisely that of the extremal Reissner-Nordstr\"om
solution.   In the second case, the divergence of ${\cal M}/r$ at
$r=0$ ensures that there is always some small value of $r$ at which
there is a horizon.  However, the discussion above suggests that we
should not expect to find a solution with two horizons.  Hence, the
solutions of this type should disappear when  $v$ becomes large enough
that the peak at $r \sim 1/(ev)$ reaches $1/(2G)$.  But, at least for
small values of ${\cal M}(0)$, the behavior near this peak should be
just as for the first case, and so the maximum mass here also should
be simply the  extremal Reissner-Nordstr\"om mass.    Finally, for the
third case the plot would suggest that there is always a horizon.
However, as was argued above, we should not expect to find a solution
with nontrivial fields outside the horizon if $r_H \ga R_{\rm mon}$.
If $r_H \approx {\cal M}(0)\, G$, this means that new black hole
solutions should exist for this case only if $M ={\cal M}(\infty)
\approx {\cal M}(0) \la M_{\rm ext}^2/M_{\rm mon}$.

\bigskip

\centerline{\epsffile{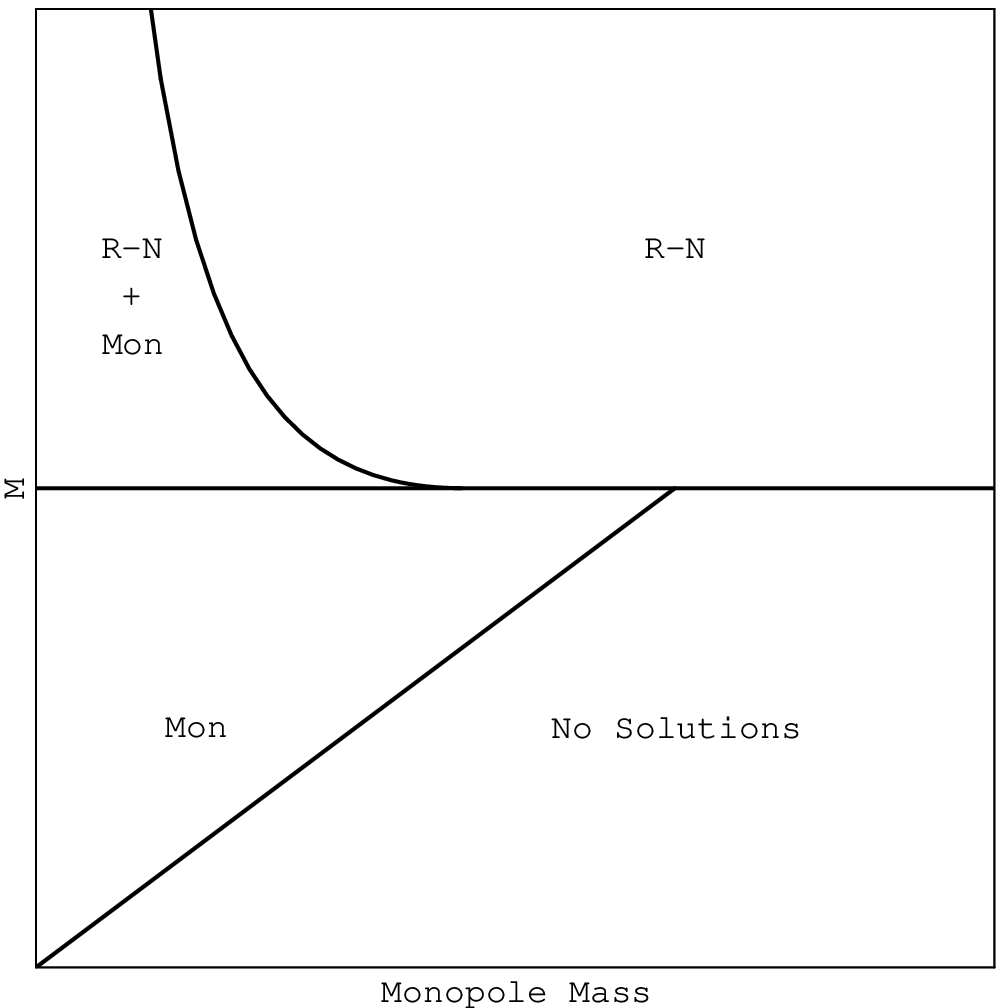}}
\medskip
\centerline{\vbox{ \tenrm \baselineskip=12truept  \hsize=4.0truein
\noindent Fig.~3.  Schematic phase diagram of solutions. ``R-N'' refers to a
Reissner-Nordstrom solution with a horizon, while ``Mon'' refers to
the solutions with a black hole inside a nontrivial monopole
configuration. }}

      These results are summarized by the phase diagram in Fig.~3,
which indicates the types of magnetic black holes that exist for various
values of the total mass $M$ and the flat space monopole mass $M_{\rm
mon}$.   These new solutions, which may be described as black holes inside
magnetic monopoles, occur in the region on the left side of the plot; the
diagonal line which bounds this region from below corresponds to the
nonsingular magnetic monopoles.   In addition to these solutions, there
are also the purely Reissner-Nordstr\"om black holes for all $M \ge M_{\rm
ext}$; these are of course independent of the value of $M_{\rm mon}$.
Examining this plot, one sees that there is a region
where both types of black holes are allowed.  Thus, there can be two
different black hole solutions with the same mass and the same magnetic
charge.   Can one of these be transformed into the other?

   The second law of black hole dynamics prevents the black hole with hair
from  going over into the Reissner-Nordstr\"om solution (at least
classically), since for a given mass the latter has the smaller area.
On the other hand, there is no
general principle forbidding the reverse transition.  We can pursue
further the question of whether it actually occurs by studying the
stability of the Reissner-Nordstr\"om solution under small fluctuations.

\sub{4.3}{Instabilities of the Reissner-Nordstr\"om Black Hole --- A
First Examination}

   For the moment, let us consider only spherically symmetric
fluctuations
about the Reissner-Nordstr\"om solution with unit magnetic
charge\ref{K.~Lee, V.P.~Nair and E.J.~Weinberg,
\prl{68}{1100}92.}.  If growing modes can be found among these, then the
Reissner-Nordstr\"om solution is certainly unstable, although the converse
need not be the case.  After having developed the necessary formalism, I
will return later to the general stability analysis.

    The spherically symmetric fluctuations are
completely described by four functions: the deviations of the metric
functions $A(r)$ and $B(r)$ from their Reissner-Nordstr\"om form, the
deviation of the Higgs field function $h(r)$ from unity, and the function
$u(r)$ that measures the magnitude of the massive vector field and that
vanishes for the Reissner-Nordstr\"om black hole.    When the field
equations are linearized about the Reissner-Nordstr\"om solution, they
separate into three decoupled sets of equations.  One, involving
$\delta A(r,t)$ and $\delta B(r,t)$, is the same as for the purely
electromagnetic case, where it was shown some time ago\ref{V.~Moncrief,
\pr{9}{2707}74;
 \pr{10}{1057}74; \pr{12}{1526}75.} that
there is no instability.  One can easily show that the second, involving
only $\delta h(r,t)$, also does not lead to any instability.   Finally,
there is the equation for $u(r,t)$.  The generalization of Eq.~\ueq\ to
the time-dependent case is
$$ {1\over \sqrt{AB} } {\partial \over \partial t}
   \left( {\sqrt{AB} \over B} \dot u \right)
   -{1\over \sqrt{AB} } {\partial \over \partial r}
   \left( {\sqrt{AB} \over A} u' \right)
    + \left( {(u^2-1)\over r^2} -e^2v^2h^2 \right)u =0 \, .
   \eqno\eq $$
By setting $A$ and $B$ equal to their Reissner-Nordstr\"om values and $h=1$,
and keeping terms linear in $u$, we obtain
$$ -{1\over B_{RN}} \ddot u - {\partial \over \partial r} (B_{RN} u')
     + \left( {1\over r^2} -e^2v^2 \right) u =0 \, .
    \eqn\firsteq $$
Any instability will appear as an exponentially growing solution of
this equation; i.e., one of the form
$$ u(r,t)= u(r)e^{\omega t} \, .
   \eqn\expu $$
Let us define  a tortoise coordinate $x$ by
$$  {dr \over dx} = B_{RN}(r) \, .
    \eqno\eq $$
(Note that the range $r_H \le r < \infty$ corresponds to $-\infty < x <
\infty$.)
Making this change of variables in Eq.~\firsteq\ and requiring $u$ to
behave as in Eq.~\expu, we obtain
$$  -{d^2u \over dx^2} + {B_{RN}(r)\over r^2} (e^2v^2 r^2 -1) u
    = -\omega^2 u  \, .
    \eqn\schrodeq $$
This is of the form of the Schr\"odinger equation for a particle in the
potential
$$  V(x) =  {B_{RN}(r)\over r^2} (e^2v^2 r^2 -1) \, .
   \eqno\eq $$
The existence of an unstable mode about the Reissner-Nordstr\"om
solution is equivalent to the existence of a negative energy bound
state in the potential $V(x)$.  If $r_H > 1/(ev)$, this potential is
positive over the whole range of $x$, and so there are clearly no bound
states.  On the other hand, as $r_H$ falls below $1/(ev)$ the potential
becomes negative near the horizon, raising the possibility of a bound
state.  Numerical calculations\Ref\Alex{S.A.~Ridgway and
E.J.~Weinberg, \pr{51}{638}95.}\ reveal
that a bound state is present if $$   r_H < r_{\rm cr} \approx {0.557
\over ev} \, .
    \eqno \eq $$
Thus, the Reissner-Nordstr\"om black hole becomes unstable when its horizon
is of the order of the monopole radius.  In terms of mass, the
instability occurs whenever the black hole mass $M < M_{\rm unstable}
\approx M^2_{\rm Pl}/(ev)$. This is greater
than the mass of the extremal Reissner-Nordstr\"om solution as long as $v <
M_{\rm Pl}$, i.e., for essentially  any value of $v$ consistent with the
existence of nonsingular monopoles.

      We can now use these results to follow the evolution in such a
theory of an initially large black hole carrying a unit magnetic charge,
such as might have formed from the collapse of star containing  a
single magnetic monopole.  (The multiply-charged case will be described
later.) After the collapse, the black hole
eventually settles down to the Reissner-Nordstr\"om form, and slowly
evaporates via the Hawking process.  If the black hole were to remain
Reissner-Nordstr\"om, this evaporation would cease as it reached the
extremal solution with vanishing Hawking temperature.  Instead, as the
evaporation causes the horizon to fall below the monopole radius, the
core of an 't~Hooft-Polyakov monopole begins to form outside the
horizon.  As the evaporation continues, the receding horizon reveals
more of the monopole core, while the Hawking temperature (which can be
obtained from Eq.~\Htemp) rises
monotonically.   This continues until the horizon reaches Planck size,
at which point the black hole presumably evaporates completely, leaving
behind a nonsingular monopole.

\main{5. Nontopological Magnetic Monopoles and New Magnetic Black Holes}
\nobreak

     The discussion of magnetic monopoles in the previous two
sections was framed largely in context of a spontaneously broken
gauge theory in which monopoles appear as classical topological
solitons.  However, one can construct a wider class of theories which
have classical solutions carrying magnetic charge even though there is
no nontrivial topology\Ref\kml{K.~Lee and
E.J.~Weinberg, \prl{73}{1203}94.}.  Imposing a certain
constraint on the parameters ensures that the energy density is
nonsingular at the origin.  This condition can be relaxed  if one is
interested in black hole solutions, where the singularity can be hidden
within the black hole horizon.

     Thus, consider a theory with electromagnetism, a massive
charged spin-1 particle represented by a complex vector field
$W_\mu$, and a neutral massive scalar $\phi$.  The Lagrangian is
$$ \eqalign{ {\cal L} &=  -{1\over 2}|D_\mu W_\nu
- D_\nu W_\mu|^2
 -{1\over 4} F_{\mu\nu}F^{\mu\nu} + {g\over 4} d_{\mu\nu} F^{\mu\nu}
  -{\lambda\over 4} d_{\mu\nu} d^{\mu\nu} \cr
  &\qquad
  + m^2(\phi) |W_\mu|^2 + {1\over 2} \partial_\mu \phi \,
\partial^\mu\phi
   - V(\phi) \cr}
    \eqn\generallagrangian$$
where
$$ F_{\mu\nu} = \partial_\mu A_\nu - \partial_\nu A_\mu
     \eqno \eq$$
and
$$   D_\mu = \partial_\mu -ieA_\mu
    \eqno \eq $$
are the ordinary electromagnetic field strength and covariant
derivative, respectively, and
$$ d_{\mu\nu} = ie (W_\mu^* W_\nu -W_\nu^* W_\mu) \, .
   \eqno\eq $$
The term involving $d_{\mu\nu}F^{\mu\nu}$ gives an anomalous magnetic
moment to the vector particles that combines with the
contribution from the covariant derivative terms to give a total
gyromagnetic ratio $(g+2)/2$.  The quantity $m(\phi)$ represents a
$\phi$-dependent mass for the $W$; let us denote by $m_W$ the value that
this takes when $\phi$ is chosen to minimize the potential $V(\phi)$.

      If $g=2$, $\lambda=1$, and
$m(\phi)=e\phi$, this theory is simply the $SU(2)$ gauge theory of
Eq.~\Higgslag\ written in the unitary gauge where $\phi^a$ has a single nonzero
component, $\phi^3 \equiv \phi$.  The vectors fields in
Eqs.~\Higgslag\
and \generallagrangian\
are related by $A_\mu^3 = A_\mu$ and $(A_\mu^1 + i A_\mu^2)/\sqrt{2} =
W_\mu$.  Similarly, for  $g=2$, $\lambda= (\sin \theta_W)^{-2}$,
and $m(\phi)= e\phi/2$, Eq.~\generallagrangian\  gives the unitary gauge
form of the standard electroweak theory, but with all terms involving the
$Z$ or fermions omitted.  For generic values of the parameters, however,
the extension to the standard non-Abelian theory is not possible.  In
such cases $\phi$ cannot be viewed as a component of a Higgs multiplet
and the usual topological arguments for the existence of a solution
cannot be applied.  Nevertheless, energetic considerations show that
magnetically charged solutions  exist.  (An extension to a somewhat
nonstandard model where a topogical charge can be defined is given
in\ref{C.~Lee and P.~Yi, Seoul National University/Caltech preprint
SNUTP-94-73/CALT-68-1944 (1994).}.)

      To see this, let us begin with a configuration where $W_\mu$ is
identically zero, $\phi$ is everywhere equal to its vacuum value,
and the electromagnetic vector
potential $A_i(x)$ is given by the Dirac potential of
Eq.~\Diracpotential\  with $Q=1/e$.  This
carries unit magnetic charge, but has infinite energy because of the
divergent Coulomb magnetic field near the origin.   We can reduce, and
possibly even remove, this divergence by surrounding this singularity
with a cloud of $W$-particles with their magnetic moments oriented so as
to cancel the original magnetic field.  To examine this in further
detail, we need the explicit expression for the energy density of a
static configuration.   For the sake of simplicity, let us assume that
the configuration carries no electric charge, and that $A_0=W_0=0$.  The
energy density is then
$$ \eqalign{ {\cal E} &=
   {1\over 4}\left( 1- {g^2 \over 4\lambda}\right) F_{ij}^2
   + {g^2 \over 16\lambda}
     \left( F_{ij} - {2\lambda \over g} d_{ij} \right)^2
    +{1\over 2} |D_iW_j - D_jW_i|^2
   \cr &\qquad + m^2(\phi) |W_i|^2  +{1\over 2} (\partial_i \phi)^2
        +  V(\phi) \,.}
   \eqn\energy $$

  The first term on the right hand side behaves as $1/r^4$ near the
origin; it represents that part of the Coulomb energy that cannot be
cancelled by the $W$ magnetic moments.  Let us assume for the moment that
$g^2=4\lambda$, so that this term is absent.   The next term can be made
finite by choosing $W_i$ in such a fashion that $d_{ij}$
has a $1/r^2$ singularity at the origin
that cancels the singularity in $F_{ij}$.
This requires that the magnitude of $W_i$ grow as $1/r$ near the
origin, which would lead one to expect $1/r^4$ contribution to the energy
density from  the angular derivatives in the next term.   However, it
turns out to be possible to arrange the vector field so that these angular
derivatives cancel completely, leaving only radial derivative terms that
are easily made finite.  (The discussion in the next section will
clarify this.)  Next is the mass term for the $W$.
With $W_i \sim 1/r$, this term would have an integrable $1/r^2$
singularity; this singularity can be eliminated completely by choosing
$\phi(x)$ so that $m(\phi)$ vanishes at the origin.   Finally, the last
two terms, involving only the scalar field, cause no difficulty.

     We can thus obtain a finite energy configuration carrying unit
magnetic charge.  There is a Dirac string singularity, but we know that
this is not physically observable.  There may also be singularities in the
vector fields at the origin, but these do not lead to any singularity in
the energy density.   With parameters chosen so the theory is actually
the unitary gauge form of the $SU(2)$ Higgs theory, this configuration is
simply the unitary gauge form  of the 't~Hooft-Polyakov monopole given in
Eq.~\unitaryform.  For other choices of parameters, but with $g^2=4\lambda$, it
describes a nontopological, but finite energy, magnetic monopole.

      Now consider relaxing the condition $g^2=4\lambda$.  If $g^2 >
4\lambda$, the first term in Eq.~\energy\ leads to an infinite negative
energy at the center of any monopole configuration, implying that the
vacuum is unstable against formation of monopole-antimonopole pairs.  If
instead $g^2 < 4\lambda$, any monopoles will have infinite energy.
Although this rules out the existence of such objects in flat spacetime,
it leaves the possibility of magnetic black holes in which this
singularity is hidden behind the horizon.  We can pursue this possibility
using methods similar to those of the previous section.   Thus, let us
assume spherical symmetry, with the metric being of the form of
Eq.~\sphericalmetric\ and the vector field determined by a single
function $u(r)$, as before, while the scalar field is a function only of
$r$.  The matter action is again of the form of Eq.~\matteraction, with
$$ K = {{u^\prime}^2 \over e^2r^2} + {1\over 2} {\phi^\prime}^2
     \eqno\eq$$
and
$$  U = {\lambda\over 2e^2r^4}
    \left( u^2- {g\over 2\lambda} \right)^2
        + {u^2 m^2(\phi) \over r^2 } + V(\phi)
        +   {1\over 2e^2r^4}\left(1 - {g^2\over 4\lambda}\right) \, .
     \eqno\eq$$
If we define a function $F(r)$ by
$$ A(r) = \left[ 1 -{2G {F}(r)\over r} +{4\pi G\over r^2 e^2}
     \left( 1-{ g^2 \over 4\lambda} \right) \right]^{-1} \, ,
     \eqn\Ainv $$
the gravitational field equations imply that
$$ F^\prime = 4\pi r^2 \left( {K\over A} + U_1  \right) \,.
      \eqn\Meq$$

     The horizon, $r_H$, is a zero of $A^{-1}(r)$.  The existence of such
a zero places a lower bound
$$   F(r_H) \ge {\sqrt{4\pi}\over e} m_{\rm Pl}
    \left( 1-{ g^2 \over 4\lambda} \right)^{1/2}
  = M^{RN}_{\rm ext} \left( 1-{ g^2 \over 4\lambda} \right)^{1/2}
     \eqno\eq  $$
on the value of $F(r)$ at the horizon, and implies a lower bound on the
total mass $M = F(\infty) \approx F(r_H) + m_W/e^2$, where the second
term is an estimate of the sum of the energy in the portion of
the monopole core outside the horizon and that contained in the
long-range Coulomb field.  A rough upper bound on the mass can be obtained
by requiring that the horizon radius be less than that of the monopole
core.  Together, these considerations lead us to expect new magnetic
black hole solutions with masses in the range
$$   \left( 1-{ g^2 \over 4\lambda} \right)^{1/2}   M^{RN}_{\rm ext}
       \la\, M\, \la \,
      {\sqrt{g} \, m_{\rm Pl}^2 \over  m_W}  \, ,
   \eqno\eq $$
where it has been assumed that $m_W \ll \mpl$.
The solutions saturating the lower bound will be new types of extremal
black holes.  Since they have a lower mass, but the same magnetic charge,
as the extremal Reissner-Nordstr\"om solutions, there will be a repulsive
force between any two extremal holes with magnetic charges of the same
sign.

      We can also envision solutions with other values of the magnetic
charge, subject to the constraint that $q = eQ_M$  be either an integer
or an integer plus one-half.  (This quantization condition can be obtained
classically by requiring that the Dirac string not affect the scattering
of classical $W$-field waves.)   These solutions cannot be spherically
symmetric and so require a more complicated analysis than those with
$q=1$.  The first step toward dealing with these is to  develop the
formalism of monopole spherical harmonics; this is done in the next
section.

\main{6. Angular Momentum and Monopole Spherical Harmonics}
\nobreak

    It is well known that in the presence of a magnetic monopole an
electrically charged particle acquires an additional angular
momentum oriented toward the position of the monopole and with a
magnitude equal to the product of the electric and magnetic charges.
This extra term changes the spectrum of eigenvalues of the orbital
angular momentum in the quantum theory, and leads to a modification of
the corresponding eigenfunctions, the spherical harmonics.

    Thus, a spinless particle with electric charge $e$ moving in the
field of a monopole with magnetic charge $q/e$ has a conserved orbital
angular momentum
$$  \eqalign{{\bf L} &= {\bf r \times } (m{\bf v}) - q {\bf \hat r} \cr
    &= {\bf r \times} ( {\bf p} - e{\bf A} ) - q {\bf \hat r} \, .}
   \eqn\monoangmom $$
This can be represented in the usual fashion by the differential operator
$$  {\bf L } = -i {\bf r \times D} - q {\bf \hat r}
    \eqno\eq $$
where
$$    {\bf D} = {\bf \nabla} -ie {\bf A} \, .
    \eqno\eq $$
The components of $\bf L$ obey the standard angular momentum commutation
relations.  The monopole spherical
harmonics\REFS\Tamm{I.~Tamm, {\it Z.~Phys.}~{\bf 71}, 141
(1931).}\REFSCON\WuYang{T.T.~Wu and C.N.~Yang, \np{107}{365}76.}\refsend\
$Y_{qlm}(\theta,\phi)$ are
eigenfunctions of ${\bf L}^2$ and $L_z$ obeying
$$ \eqalign{   {\bf L}^2 Y_{qlm}(\theta,\phi)
   &= l(l+1) Y_{qlm}(\theta,\phi) \cr
    L_z Y_{qlm}(\theta,\phi) & = m Y_{qlm}(\theta,\phi) \, .}
    \eqno\eq $$
The eigenvalues $m$ run in integer steps from $l$ down to $-l$, as
usual.   The allowed values of $l$, however, are not the usual
ones.   Classically, the fact that the two terms on the right hand side
of Eq.~\monoangmom\ are perpendicular to each other implies that  ${\bf
L}^2 \ge q^2$.  Correspondingly, the minimum value of $l$ is not 0, but
rather $q$, with all integer increments above this also allowed.

   As with the usual spherical harmonics, the monopole harmonics for a
given value of $q$ form a complete and orthonormal set, with
$$  \int d\Omega\, Y^*_{qlm}(\theta,\phi) Y_{ql'm'}(\theta,\phi)
     = \delta_{ll'} \delta_{mm'} \, .
   \eqno\eq $$

     The precise form of the monopole harmonics is gauge-dependent,
reflecting the gauge dependence of the vector potential.   In addition,
they have singularities corresponding to the Dirac string singularity of
the potential.  In a gauge where
$$   A_\phi = {q\over e} (\pm 1 -\cos \theta)
  \eqn\stdpotential $$
the monopole harmonics are of the form
$$  Y_{qlm}(\theta,\phi)  = e^{i(m\pm q)\phi} F_{qlm}(\theta) \, .
   \eqno\eq $$
For example, with $q =1/2$,
$$   \eqalign { F_{{1\over 2} {1\over 2}{1\over 2}} &=
   -{1\over \sqrt{4 \pi}} \sqrt{1-\cos\theta} \cr
    F_{{1\over 2} {1\over 2}-{1\over 2}} &=
   -{1\over \sqrt{4 \pi}} \sqrt{1+\cos\theta}\, . }
   \eqno\eq $$
With the upper choice of sign, $Y_{1/2, 1/2, 1/2}$ is singular along the
negative $z$-axis, where the Dirac string is located.  If one chooses
instead the lower sign, then these singularities are replaced by a Dirac
string along the positive $z$-axis, with a corresponding singularity in
$Y_{1/2, 1/2, -1/2}$.  As emphasized by Wu and Yang\refmark{\WuYang}, one
can combine these, using one gauge for the region $0 \le \theta < (\pi/2)
+ \delta$ and the other for $ (\pi/2) -\delta < \theta \le \pi$, to obtain
an object that is nonsingular everywhere but at the origin.

     For dealing with charged vector fields we will need monopole
vector spherical harmonics, which are eigenfunctions of the total
angular momentum ${\bf J} = {\bf L} + {\bf S} $.  By the usual rules for
adding angular momenta, one sees that the minimum value for the total
angular momentum quantum number $J$ is $q-1$, except in the two cases
$q=0$ and $q=1/2$, where $J_{\rm min}=q$.  Note that $J=0$ is possible
only if $q=0$ of $q=1$; this is why  nonsingular spherically symmetric
monopoles can only have unit magnetic charge in the spontaneously broken
$SU(2)$ theory\ref{A.H.~Guth and E.J.~Weinberg, \pr{14}{1660}76.}.

    In general, there are several multiplets of monopole harmonics
with a given value of $J$.   These can be  chosen to be
eigenfunctions of ${\bf L}^2$\ref{H.A.~Olsen, P.~Osland, and T.T.~Wu,
\pr{42}{665}90.}.  However,
for our purposes it will be more convenient to choose
harmonics\Ref\harmonic{E.J.~Weinberg, \pr{49}{1086}94.}\  that are
eigenfunctions of
$$  \lambda \equiv {\bf \hat r \cdot S} \, .
    \eqno\eq $$
In general, the eigenvalues of $\lambda$ are $-1$, 0, and 1.  However,
the identity
$$  {\bf \hat r \cdot J} = {\bf \hat r \cdot L} + {\bf \hat r \cdot S}
      = -q + \lambda
   \eqno\eq $$
leads to the restriction
$$  -J \le \lambda - q \le J \, .
    \eqno\eq $$
{}From this, one sees that the multiplet structure is as follows:
\item{} For $J= q-1 \ge 0$, one multiplet, with $\lambda =1$.
\item{} For $J= q  > 0$, two multiplets, with $\lambda =1$ and 0.
\item{} For $J= q = 0$, one multiplet, with $\lambda =0$.
\item{} For $J> q$, three multiplets, with $\lambda =1$, 0, and $-1$.

 Thus, let us define monopole vector harmonics ${\bf C}^{(\lambda)}_{qJM}$
that obey
$$  \eqalign{ {\bf J}^2 {\bf C}^{(\lambda)}_{qJM}
   &= J(J+1) {\bf C}^{(\lambda)}_{qJM} \cr
    J_z {\bf C}^{(\lambda)}_{qJM} &= M {\bf C}^{(\lambda)}_{qJM} \cr
     {\bf \hat r \cdot S}  {\bf C}^{(\lambda)}_{qJM}
          &= \lambda {\bf C}^{(\lambda)}_{qJM} \, . }
   \eqno\eq $$
Because the spin matrices $(S^k)_{ij} = -i\epsilon_{ijk}$, the last of
these can be rewritten as
$$  {\bf \hat r} \times {\bf C}^{(\lambda)}_{qJM}
     =  -i \lambda {\bf C}^{(\lambda)}_{qJM}  \, .
    \eqn\rcross $$
{}From this we see that the harmonics with $\lambda=0$ are purely radial,
while those with $\lambda = \pm 1$ have only angular components.
Further, we find that harmonics with different
values of $\lambda$ are orthogonal vectors at every point, in the sense
that
$$  {\bf C}^{(\lambda)*}_{qJM} \cdot {\bf C}^{(\lambda')}_{qJ'M'}
   =0,  \qquad \lambda \ne \lambda' \, .
    \eqno \eq $$
Harmonics with different values of $j$, $M$, or $\lambda$ are also
orthogonal in the functional sense.  Using a convenient choice of
normalization, we have
$$   \int d \Omega \,{\bf C}^{(\lambda)*}_{qJM}
      \cdot {\bf C}^{(\lambda')}_{qJ'M'}
   = { \delta_{JJ'} \delta_{MM'} \delta_{\lambda\lambda'} \over r^2}
    \eqn\vecharmnorm $$

      Experience with the standard vector harmonics would suggest that
the monopole spherical harmonics could be constructed by applying vector
differential operators to the scalar monopole harmonics.  This
construction does in fact work if $J \ge q$, although not for the case $J
=q-1$, which will be treated separately below.  Let $\bf v$ be any
vector constructed from $\bf r$ and $\bf D$.  It follows from the
commutation relations
$$  \eqalign{ [{\bf L}^2, v_k] &= - 2 i \epsilon_{ijk} L_iv_j - 2v_k \cr
    &= -2 \left[ ({\bf L \cdot S} + 1 ) {\bf v} \right]_k }
   \eqno \eq $$
that
$$  ({\bf L} + {\bf S} )^2 {\bf v} Y_{qKM} = {\bf v}{\bf L}^2 Y_{qKM}
       = K(K+1) {\bf v} Y_{qKM}  \, .
     \eqno \eq $$
Hence, if ${\bf v}_\lambda$ is a vector satisfying ${\bf \hat r}\times
{\bf v}_\lambda = - i\lambda {\bf v}_\lambda$,  the desired harmonics
are, up to a normalization constant, equal to ${\bf v}_\lambda Y_{qJM}$.
Explicitly,
$$ \eqalign{  {\bf C}^{(1)}_{qJM} &= {1 \over \sqrt{2( {\cal J}^2 +q) }}
    \left[ {\bf D} + i {\bf \hat r} \times {\bf D} \right] Y_{qJM},
    \qquad J \ge q > 0  \cr
        {\bf C}^{(0)}_{qJM} &=  { {\bf \hat r} \over r} Y_{qJM},
  \qquad J \ge q \ge 0  \cr
     {\bf C}^{(-1)}_{qJM} &= {1 \over \sqrt{2( {\cal J}^2 -q) }}
    \left[ {\bf D} - i {\bf \hat r} \times {\bf D} \right] Y_{qJM},
    \qquad J >q \ge 0 }
  \eqno\eq $$
where
$$  {\cal J}^2 \equiv J(J+1) - q^2  \, .
   \eqno \eq $$

    Formulas for the covariant divergence and covariant curl of these
vector harmonics are easily derived.  Consider first the $\lambda =0$
harmonics.  For their curl, we have
$$  \eqalign {{\bf D} \times {\bf C}^{(0)}_{qJM}
   &= {\bf D} \times { {\bf \hat r} \over r} Y_{qJM}
   = - { {\bf \hat r} \over r} \times {\bf D} Y_{qJM} \cr
    &= {i\over r} \left[
   \sqrt{{\cal J}^2 + q\over 2} {\bf C}^{(1)}_{qJM}
  - \sqrt{{\cal J}^2 - q\over 2} {\bf C}^{(-1)}_{qJM} \right] \, ,}
    \eqno \eq $$
while for their divergence,
$$ {\bf D \cdot} {\bf C}^{(0)}_{qJM}
  = {\bf D \cdot} \left({ {\bf \hat r} \over r} Y_{qJM} \right)
   = {1 \over r^2 } Y_{qJM} \, .
    \eqno \eq $$
For the $\lambda =\pm 1$ harmonics, the first step is to note the identity
$$  {\bf  r}\times \left( {\bf D} \times
   {\bf C}^{(\pm 1)}_{qJM}  \right)
   = {\bf D} \left({\bf r \cdot}  {\bf C}^{(\pm 1)}_{qJM} \right)
 - {\bf C}^{(\pm 1)}_{qJM} - ({\bf r \cdot D}) {\bf C}^{(\pm 1)}_{qJM}\, .
    \eqn\covcurl $$
The first term on the right vanishes because the $\lambda =\pm 1$
harmonics have
only angular components, while the last two cancel because the harmonics
are homogeneous of degree $-1$ in the Cartesian coordinates.  Hence the
entire right hand side vanishes, implying that ${\bf D} \times
{\bf C}^{(\pm 1)}_{qJM} $ is purely radial and thus a linear combination
of the $\lambda=0$ harmonics.  The coefficients can be obtained from
$$  \int d\Omega \,{\bf C}^{(0)*}_{qJ'M'} \cdot {\bf D} \times
   {\bf C}^{(\pm 1)}_{qJM} =
    -\int d\Omega \,{\bf D} \times {\bf C}^{(0)*}_{qJ'M'} \cdot
      {\bf C}^{(\pm 1)}_{qJM}  \, .
    \eqno \eq $$
Using the result already obtained for the curl of ${\bf C}^{(0)}_{qJM}$,
together with the normalization condition \vecharmnorm, one then obtains
$$   {\bf D}\times  {\bf C}^{(\pm 1)}_{qJM} = \pm {i\over r}
     \sqrt{{\cal J}^2 \pm q\over 2} {\bf C}^{(0)}_{qJM}\, .
    \eqno \eq $$
Finally, for the covariant divergence we have
$$  \eqalign{ {\bf D \cdot} {\bf C}^{(\pm 1)}_{qJM} &=
     \pm i {\bf D \cdot}
     \left( {\bf \hat r} \times {\bf C}^{(\pm 1)}_{qJM}  \right) \cr
   &= \mp i {\bf \hat r} \cdot{\bf D} \times {\bf C}^{(\pm 1)}_{qJM} \cr
    &= {1\over r^2} \sqrt{{\cal J}^2 \pm q\over 2}  Y_{qJM} \, . }
    \eqn\covdiv $$

    The vector harmonics with $J=q-1$ form a single multiplet, with $\lambda
=1$.  They clearly cannot be obtained by a construction of the above type,
since there are no scalar harmonics with $J=q-1$.   To begin, I will show
that both their covariant curls and their covariant divergences
vanish.  (This gives the cancellation of angular derivatives that was
asserted in the discussion following Eq.~\energy.) To
simplify the notation, let
$$   {\bf U}_M = {\bf C}^{( 1)}_{q,q-1,M} \, .
      \eqno\eq $$
First consider the divergence ${\bf D\cdot U}_M$.  Since this is a
scalar, it can be expanded as a linear combination of the scalar
monopole harmonics, with the coefficient of $Y_{qJ'M'}$ being
$$  \eqalign{ I_{J'M'} &= \int d\Omega\, Y_{qJ'M'}^* {\bf D\cdot U}_M \cr
    &  = - \int d\Omega \,({\bf D}Y_{qJ'M'}^*) \cdot {\bf U}_M  \cr
    &= -\int d\Omega \,\left[ \sqrt{{\cal J}^{'2} + q\over 2}
              {\bf C}^{(1)*}_{qJ'M'}   +  \sqrt{{\cal J}^{'2} - q\over 2}
              {\bf C}^{(-1)*}_{qJ'M'} \right] \cdot {\bf U}_M \, . }
    \eqno\eq $$
The last integral is zero, since the harmonics with  $J' \ge q$ are
orthogonal to those with $J=q-1$.   Since all of the $I_{J'M'}$ vanish, so
must  ${\bf D\cdot U}_M$.  Next, manipulations paralleling those in
Eq.~\covcurl\
show that the angular components of ${\bf D} \times {\bf U}_M $ vanish.
Finally, the analog of Eq.~\covdiv\ shows that the radial component of the
curl is proportional to the divergence, which has just been shown to be
zero.  Hence,  ${\bf D} \times {\bf U}_M =0$.
The $J=q-1$ vector harmonics can therefore be viewed as a set of
curl-free and divergenceless vector fields on the unit two-sphere.   One can
derive an index theorem to the effect that there are precisely $2q-1 =
2(q-1)+1$ linearly independent fields of this sort on the two-sphere,
giving just the number needed for a $J=q-1$ multiplet.

     Let us construct these explicitly, working in a gauge where the vector
potential is given by Eq.~\stdpotential.  When acting on a scalar quantity,
$J_z = L_z = -i\partial_\phi \mp q$.  Let us apply this to the scalar
${\bf \hat z \cdot U}_M$, where $\bf \hat z$ denotes the unit vector along
the $z$-axis.  Using the fact that $\bf \hat z$ is invariant under
rotations about the $z$-axis and therefore commutes with $J_z$, we have
$$ J_z ({\bf \hat z \cdot U}_M ) = M ({\bf \hat z \cdot U}_M )
   = (-i\partial_\phi \mp q) ({\bf \hat z \cdot U}_M )
     \eqno \eq $$
from which it follows that
$$  (-i\partial_\phi \mp q) ({\bf U}_M)_\theta  = M ({\bf U}_M)_\theta\, .
    \eqno\eq $$
Hence, the $\theta$-components of these harmonics must be of the form
$$  ({\bf U}_M)_\theta = e^{i(M \pm q)\phi} f_{qM}(\theta) \, .
   \eqno \eq $$
Next, applying Eq.~\rcross\ to the the $J=q-1$ harmonics gives
$$  ({\bf U}_M)_\phi = i \sin\theta ({\bf U}_M)_\theta \, .
    \eqno \eq $$
This identity, together with the vanishing of the curl, leads to the
differential equation
$$  \partial_\theta( \sin\theta f_{qM}) - (M+\cos\theta) f_{qM} =0 \, .
       \eqno\eq $$
Solving this for the $f_{qM}$, we obtain
$$  \eqalign {({\bf U}_M)_\theta
  &= a_{qM}  e^{i(M \pm q)\phi} (1-\cos\theta)^M (\sin\theta)^{q-M+1} \cr
 &= a_{qM}  e^{i(M \pm q)\phi} (1+\cos\theta)^M (\sin\theta)^{q+M-1} \, .}
      \eqno \eq $$
{}From the form given on the first line it is clear that the harmonic
is singular along the positive $z$-axis unless $q-M+1
\ge 0$.  Similarly, the form on the second line
is manifestly singular along the negative $z$-axis unless $q+M-1 \ge 0$.
To satisfy both constraints, $M$ must lie in the interval $-(q-1) \le M
\le q-1$.

\main{7. Stability Analysis of the Reissner-Nordstr\"om Solution}
\nobreak

      Armed with the results of the previous section, we are now
prepared to carry out the full stability analysis\refmark{\Alex}\ for a
Reissner-Nordstr\"om black hole with arbitrary magnetic charge in the more
general  theory described by the matter Lagrangian~\generallagrangian.
This will include as a special case the analysis in Sec.~4.3 of the
spherically  symmetric perturbations about a singly-charged black hole in
the context of the Higgs theory.  Just
as in that case, the fluctuations of the massive vector fields separate
from those of the metric and of the electromagnetic field; since it is
known that the latter two do not give rise to any instability, we can
restrict our attention to the fluctuations in $W_\mu$.   Keeping only
terms in the matter action that are quadratic in $W_\mu$, we obtain
$$\eqalign{ S_{\rm quad} = \int d^4x \sqrt{-\det(g_{\mu\nu})} \Biggl[
        & -{1\over 2} \left| D_\mu W_\nu - D_\nu W_\mu \right|^2
        - m^2 \left| W_\mu \right|^2 \qquad\cr
   &\qquad -{ieg \over 4} F^{\mu\nu} \left( W^*_\mu W_\nu - W^*_\nu W_\mu
    \right) \Biggr] . }
       \eqn\quadaction $$
Here, and for the remainder of this section,
$m(\phi)$ should be understood to be at its vacuum value $m_W$.

    We can exploit the spherical symmetry of the unperturbed solution by
expanding $W_\mu$ in terms of scalar and vector spherical harmonics:
$$ \eqalign{ W_t &= \sum_{J=q}^\infty \sum_{M=-J}^J
                             a^{JM}(r,t) Y_{qJM} \cr
            W_r &= {1\over r} \sum_{J=q}^\infty \sum_{M=-J}^J
                            b^{JM}(r,t) Y_{qJM} \cr
            W_a &=   \sum_{J=q-1}^\infty \sum_{M=-J}^J
                     f^{JM}_+(r,t) \left[\c^{(1)}_{qJM}\right]_a
                +    \sum_{J=q+1}^\infty \sum_{M=-J}^J
                     f^{JM}_-(r,t) \left[\c^{(-1)}_{qJM}\right]_a , }
       \eqno\eq$$
where a Roman subscript from the beginning of the alphabet stands for
$\theta$ or $\phi$.
When these expansions are substituted into the quadratic action, it
separates into a sum of terms, each of which contains only contributions
from modes with fixed values of $J$ and $M$; i.e.,
$$ S_{\rm quad}= \sum_{J=q-1}^\infty \sum_{M=-J}^J S_{\rm quad}^{JM} .
\eqno\eq$$
For $J=q-1$ the action is particularly simple, both because there is only a
single radial function, $f_+(r)$, and because the vanishing of the
covariant curl of the $J=q-1$ vector harmonics eliminates some terms.  One
finds that\footnote{2}{To simplify the notation, the superscripts $JM$
on the radial functions have been omitted in these and later equations.}
$$    \eqalign{ S_{\rm quad}^{(q-1)M}
        &= \rtint  \left\{ {1\over B} |\dot f_+|^2
     - B |f'_+|^2 - m^2 |f_+|^2 +  {qg\over 2r^2} | f_+|^2
                \right\} . }
    \eqn\lowest$$
For $J \ge q$ the action is
$$  \eqalign{ S_{\rm quad}^{JM}
         &= \rtint \left\{ | \dot b -ra' |^2
    + {1\over B}\left[\left|\dot f_+ - \splus a \right|^2
          + \left|\dot f_- - \sminus a \right|^2 \right] \right.\cr
   &\qquad
   - B\left[ \left| f'_+ - {1\over r}\splus b\right|^2
     + \left| f'_- -{1\over r} \sminus b\right|^2 \right]
     -{1\over r^2} \left| \splus f_+ - \sminus f_- \right|^2 \cr
  &\qquad \left.
    - m^2 \left[ | f_+|^2 + | f_-|^2
          + B|b|^2 - r^2{1\over B}|a|^2 \right]
     +  {gq\over 2r^2}\left[ | f_+|^2 - | f_-|^2 \right]
                \right\}  }
    \eqn\quadpertaction$$
where
$$ k_\pm = \sqrt{{\cal J}^2 \pm q \over 2}
    = \sqrt{J(J+1) - q^2 \pm q \over 2} \,\, .
     \eqno\eq $$
(There is some simplification in the case $J=q$, because the $f_-$
terms are absent.)

     Apart for the $J=q-1$ case, these are daunting expressions.  However,
some simplifications are possible.  Because its time derivative  does
not enter the action, $W_t$ (or $a^{JM}$ in the reduced actions)
is a nondynamical field that can be eliminated.  To carry this
out in detail, it is convenient to adopt a more compact notation.  For
a given $J$ and $M$, let us
combine the  radial functions entering the spatial components of $W_\mu$
into a vector $ z = \left({f_+/\sqrt{B}},\, {f_-/\sqrt{B}},\,
b\right)^T $.  The action can then be written as
$$  S_{\rm quad}^{JM}  = \rtint \left[{\dot z}^\dagger  \dot z
   + ({\dot z}^\dagger F a + a^* F^\dagger \dot z)
   + a^* G a  - z^\dagger H z \right] ,
\eqn\vectoraction$$
where
$$ F = \left( -{k_+\over \sqrt{B} }, \,-{k_-\over \sqrt{B} },\,
           r {\partial\over \partial r} \right)^T ,
\eqno \eq$$
$$ \eqalign {G &= -{\partial\over \partial r} r^2 {\partial\over \partial r}
     + {1\over B} \left( k_+^2 + k_-^2 + r^2m^2 \right) \cr
         &= \vphantom{\biggl(} F^\dagger F + B^{-1} r^2m^2 \, ,}
\eqno\eq$$
and $H$ is a $3 \times 3$ matrix whose form we do not yet need.
Variation with respect to $a^*$ and $z^\dagger$ yields the equations
$$ 0= F^\dagger \dot z + G a
    \eqno\eq $$
and
$$ 0= \ddot z + F \dot a + Hz \, ,
    \eqno\eq $$
respectively.
Using the first of these to solve for $a$ and then
substituting the result into the second equation, we obtain
$$   0= \left( I - F G^{-1} F^\dagger \right) \ddot z + Hz .
    \eqn\fieldequation$$
The coefficient of $\ddot z$,
$$  I -  F G^{-1} F^\dagger
    = I - F{1 \over F^\dagger F +B^{-1}r^2 m^2} F \, ,
    \eqno \eq $$
is a positive operator.  Using this fact, we see that unstable modes
growing as $e^{\alpha t}$ are possible if and only if the matrix $H$
has negative eigenvalues.  (However, because of the nontrivial matrix
coefficient of $\ddot z$ in Eq.~\fieldequation, the actual values of
$\alpha$ cannot
be determined directly from the negative eigenvalues.) Thus, the stability
of the Reissner-Nordstr\"om solution is equivalent to the positivity of
$H$.

    A second simplification follows from the absence of the radial
derivatives of $W_r$ from the action.  This allows us to write the
potential energy contribution associated with $H$ as
the sum of the integral of a perfect square involving $b(r)$, but not
its derivative, and a quantity that is independent of $b(r)$; i.e.,
$$  \rint \, z^\dagger H z =  E_{JM}(f_+, f_-) +
   \rint  \, B \left| k_+ f'_+ + k_- f'_-
      - {(r^2m^2 + \J^2 )\over r}\, b \right|^2   \, .
   \eqno \eq $$
By choosing $b(r)$ appropriately, the second term on the right hand side
can always be made to vanish.   Hence, the positivity of
of $H$ is equivalent to the positivity of $E_{JM}(f_+, f_-)$, which may
be written in the form
$$  E_{JM}(f_+, f_-) = \rint
    \left[ B\,{f'}^\dagger K_J f'
     + f^\dagger\left( m^2 I - {V_J \over r^2} \right) f \right] \, .
\eqn\efunc $$
where $f\equiv f_+$ if $J=q-1$ or $q$, and otherwise
$f \equiv (f_+, f_-)^T$.  For the two lowest values of $J$, the matrices
$K_J$ and $V_J$ are simply
$$   K_{q-1}= 1,  \qquad\qquad
     V_{q-1} = {gq\over 2}
   \eqno\eq$$
and
$$   K_{q}= {r^2 m^2 \over r^2m^2 + q},  \qquad\qquad
     V_{q} = {(g-2)q\over 2} \, ,   \eqno\eq$$
while for $J>q$ they are given by
$$  K_J = I - {1 \over r^2m^2 +\J^2}
      \left( \matrix{ \high k_+^2  & k_+k_- \cr
                 \high    k_+k_- & k_-^2 } \right),
    \qquad\qquad  J>q
\eqno\eq$$
$$ V_J =  \left(  \matrix{\high -k_+^2 + {qg\over 2}  &  k_+k_- \cr
                           k_+k_- & \high -k_-^2 - {qg\over 2} } \right),
      \qquad\qquad  J>q .
      \eqno\eq $$

      The eigenvalues of $K_J$ are always positive.  It follows that a
necessary condition for the existence of an instability in a mode with
angular momentum $J$  is that the potential energy terms in Eq.~\efunc\ be
negative somewhere outside the horizon.  This in turn requires that
$$   m^2 - {V_J \over r_H^2 }  < 0
    \eqn\instabcond$$
where, for $J>q$,  $V_J$ should be understood to signify the larger of its
two eigenvalues.  This condition can be phrased as an upper bound $r_H <
r_0(J)$ on the horizon radius.  For $J=q-1$, Eq.~\instabcond\ can be
satisfied only if $g>0$, in which case
$$  m r_0(q-1)  = \sqrt{ gq/2} \, .
    \eqno \eq $$
For $J=q$, instability is possible only if $g>2$, and
$$  m r_0(q)  = \sqrt{ (g-2)q/2} \, .
    \eqno \eq $$
For $J >q$, the two eigenvalues of $V_J$ are
$[ -{\cal J}^2 \pm \sqrt{{\cal J}^4 + g(g-2)q^2} ]/2$.  If $0\le g\le 2$,
both of these are negative, thus ruling out any instability.  Outside of
this range there is one positive eigenvalue, and
$$  mr_0(J) = {1\over \sqrt{2} }
   \left[ -{\cal J}^2 \pm \sqrt{{\cal J}^4 + g(g-2)q^2} \right]^{1/2} \, .
     \eqno \eq $$
Note that, for fixed $q$ and $g$, $r_0$ is a decreasing function of $J$.

     Because of the gradient terms in the energy, Eq.~\instabcond\ is not
sufficient for the existence of an unstable mode.  Instead, there is a
somewhat stronger condition that may be written as
$$  r_H < r_{\rm cr}(J) < r_0(J) \, ,
    \eqno \eq $$
where $r_{\rm cr}(J)$ is the largest value of $r_H$ for which
$$   - {d \over dr} \left[ {1\over A} K_J f' \right]
    + \left( m^2 -{V_J \over r^2} \right) f = - \omega^2 f
   \eqn\diffeq $$
has solutions with real $\omega$ that satisfy $f(r_H) =f(\infty) =0$.
(For $g=2$, $q=1$, and $J=0$, this is equivalent to  Eq.~\schrodeq.)  The
values of  $r_{\rm cr}(J)$ must be determined numerically.   To do this
for  $J=q-1$ or $q$, we first by set $\omega=0$ in Eq.~\diffeq.  We then
choose a trial value of $r_H$ and integrate in from large $r$.   By
appropriately adjusting the trial value, one can fairly quickly zero in
on the value which gives $f(r_H) =0$; this is the desired $r_{\rm
cr}(J)$.  For $J>q$, where Eq.~\diffeq\ is actually a pair of coupled
equations, this procedure cannot be applied directly.  One approach is to
replace these coupled equations by a single equation of the same order.
One such equation is obtained by replacing
the matrix $K_J$ by its smallest eigenvalue and
the matrix $V_J$ by its positive eigenvalue.  This gives an
underestimate of $E_{JM}$, and thus an overestimate of $r_{\rm cr}(J)$.
A second equation is obtained by going to a basis where $V_J$ is
diagonal and then restricting $f$ to the
subspace corresponding to the positive eigenvalue of $V_J$.  This gives
an overestimate of $E_{JM}$, and thus an underestimate of $r_{\rm
cr}(J)$.  Numerically, these two bounds on $r_{\rm cr}(J)$ turn out to
be rather close, and thus give a rather good estimate of the true value.

      Numerical calculations of $r_{\rm cr}(J)$ for a variety of
values of $q$ and $g$ indicate that $r_{\rm cr}(J)$, like $r_0(J)$, is
a decreasing function of $J$.  They also show that  $r_{\rm cr}(J)$ and
$r_0(J)$ are roughly proportional, with their ratio lying between $1/2$
and 1, except for the case $J=q-1$ with very small values of $g$, where
$r_{\rm cr}(J) \sim r^2_0(J)$.  These results indicate that as the
horizon of the Reissner-Nordstr\"om solution is decreased, instablities
occur first in the mode with the lowest possible $J$.  Thus, for $q\ge
1$, the first unstable mode has $J=q-1$ if $g>0$ and $J=q+1$ if $g<0$.
For $q=1/2$, the first unstable mode has $J=1/2$ if $g>2$ and $J=3/2$ if
$g<0$.  Finally, if $q=1/2$ and $0\le g\le 2$, the Reissner-Nordstr\"om
solution is always stable.

\main{8. Construction of New, Nonsymmetric, Black Holes}
\nobreak

      We have seen that if the horizon radius, and thus the mass, is
sufficiently small, the Reissner-Nordstr\"om black hole solution becomes
unstable.  What happens when such a black hole is perturbed
infinitesimally?   For a black hole with unit magnetic charge,
the answer is clear from the previous discussion.
The instability leads to the formation of a cloud of $W$
field outside the horizon and the black hole presumably settles
down in a new static configuration with hair.  If the matter is
described by the Higgs theory, this will be one of the new
black hole solutions described in Sec.~4.2; the extension of
these to the more general class of theories described by
Eq.~\generallagrangian\
should be straightforward.  Now consider the
case of black holes with higher charge.  If the magnetic charge of the
hole does not change, we would expect the black hole to evolve into a
new static black hole with hair, much as happens in the singly-charged
case.  However, in contrast with the previous situation, this new
solution could not be spherically symmetric, because there would be no
$J=0$ vector spherical harmonic available for the $W$-field
configuration.  Matters might be different if the black hole could
reduce its magnetic charge, either by emission of a finite energy
monopole or by fissioning into two or more holes of lower charge,
since the hole might then evolve toward one of the singly-charged
solutions we have already found.  The former possibility seems
plausible, provided that $\lambda=4g^2$ so that finite energy
monopoles exist, but energy conservation and the requirement that the
horizon area not decrease would allow this only if the initial
Reissner-Nordstr\"om black hole were well below the critical radius for
instability.  The latter possibility could occur, if at all, only
through quantum mechanical tunneling, since such splitting of a
horizon is classically forbidden.

      Thus, the instability of the Reissner-Nordstr\"om solutions leads
us to expect new black hole solutions that are not
spherically symmetric.  Without the simplifications that follow from
spherical symmetry, any attempt to obtain exact closed form
expressions for these solutions is
probably hopeless.  However, if the Reissner-Nordstr\"om is just barely
unstable (i.e., if its horizon is just slightly less than the critical
radius) a perturbative construction of these solutions is
possible\Ref\construction{S.A.~Ridgway and E.J.~Weinberg, Columbia
preprint CU-TP-673. }.

The intuition behind this construction arises from the observation
that the problem of finding a stable static solution near an unstable
one is essentially that of finding a local minimum of a function
for which one has been given a saddle point.  If the matrix of second
derivatives at a saddle point has only a single negative
eigenvalue, of sufficiently small magnitude, then we may expect to
find a local minimum near the saddle point, with the eigenvector
corresponding to the negative eigenvalue indicating the direction in
which that minimum is to be found.

To illustrate this, consider the problem of minimizing
a function of $N$ real variables $x_i$ of the form
$$ F = {1\over 2}x_i\, M_{ij}\, x_j
      + {1\over 4}\lambda_{ijkl}\,x_i \,x_j \,x_k \,x_l
    \eqno \eq $$
where the $M_{ij}$ and $\lambda_{ijkl}$ are real, and the latter are
such that the quartic term in $F$ is positive definite.  We may assume
that $M_{ij}$ is symmetric and hence can be diagonalized by an
orthogonal transformation.  Let us assume that it is diagonal in
the basis in which we are working, with diagonal matrix elements
$M_{ii} = b_i$.  The stationary points of this function are
determined by the equations
$$  0 = b_i \,x_i + \sum_{jkl} \lambda_{ijkl}\, x_j\, x_k\, x_l \, .
     \eqn\examplemin $$
If the eigenvalues of $M$ are all positive, the only
stationary point is a minimum at the origin,  $x_j=0$.  Now suppose
instead that $M$ has a single negative eigenvalue, which we may
take to be $b_1 \equiv -\beta^2$.  The minimum of $F$ no longer lies
at the origin, which is now a saddle point.  As a first step toward
the determination of this minimum, note that in the subspace defined
by  $x_1=0$ Eq.~\examplemin\ has only the trivial solution $x_j=0$.
This suggests an iterative approach to the solution in which one
first determines a nonzero approximation to $x_1$ and then uses this
to determine the remaining $x_j$.  Thus, at the first step we set
$x_j=0$ for all $j\ge 2$.  Eq.~\examplemin\ then gives the
zeroth order approximation
$$ x_1^{(0)} = \sqrt{ -{b_1 \over \lambda_{1111} }} =
          { \beta \over\sqrt{ \lambda_{1111} } } \, .
   \eqno\eq  $$
Substituting this back into Eq.~\examplemin\  leads to
$$   x_j^{(0)} = - {\lambda_{j111} \left[x_1^{(0)} \right]^3 \over b_j }
      = - {\beta^3 \lambda_{j111} \over b_j \lambda_{1111}^{3/2}}\, ,
    \qquad j\ge 2 .
    \eqno \eq $$
The next step is to substitute these lowest order approximations back
into Eq.~\examplemin\ to find higher order corrections.  For
sufficiently small $\beta$, these  higher order are
small and can be calculated as power series in $\beta$.

Let us now generalize this method to a field theory and use it to obtain
new black hole solutions.  To be specific, let us take the matter
Lagrangian to be essentially that of Eq.~\generallagrangian, but with the
$\phi$-dependent $W$ mass replaced by a constant $m$ and the remaining
scalar field terms omitted.  (This elimination of the scalar field
simplifies the calculations somewhat, but is not essential.)  The
first step is to examine the normal modes about the Reissner-Nordstr\"om
solution and identify those with negative eigenvalues.  From
previous remarks, it should be clear that the  modes involving
electromagnetic or metric perturbations all have positive
eigenvalues.  Thus, the negative eigenmodes involve only $W_\mu$ and
can be found by diagonalizing the action of Eq.~\quadaction. Specifically,
let us define ${\cal M}_{\mu\nu}$ by
$$ {\cal M}^{\mu\nu} W_\nu =
    - {1\over \sqrt{\bar g}} \bar D_\alpha \left(\sqrt{\bar g}
          W^{\alpha\mu}\right)
     +m^2 W^\mu  - {ieg \over 2}\bar F^{\alpha\mu}W_\alpha
    \eqno\eq $$
where overbars denote the use of the unperturbed metric and
electromagnetic potential.   The desired modes satisfy
$$ {\cal M}_\mu{}^\nu \psi_\nu = -\beta^2 m^2 \psi_\mu
    \eqn\eigenvalue $$
with real $\beta$; for our perturbative method to succeed, we will
need that $\beta \ll 1$.   (A factor of $m^2$ has been extracted to
make $\beta$ dimensionless.)

As in the stability analyisis, the modes can be chosen to have
definite angular momentum.  The requirements for our perturbative
method are met by choosing the horizon radius $r_H$ so that unstable
modes occur for only one value $\bar J$ of the total angular momentum,
and by furthermore requiring that $r_H$ be just less than $r_{\rm
cr}(\bar J)$; one can then show that $\beta \sim [r_{\rm cr}(\bar
J)-r_H]/r_H$.  There will then be $2\bar J +1$ degenerate negative
eigenvalues modes $\psi_\mu^M$ distinguished by the eigenvalue of
$J_z$; it is convenient to normalize these so that
$$ \int d\,^3x\sqrt{\bar g}\, \psi^{M*}_\mu\psi^{M\mu} = 1 \, ,
     \eqn\psinorm $$
where the integration is over all space outside the horizon.

We now write $W_\mu$ as a linear combination of these modes plus a
remainder, orthogonal to these, which we expect to be subdominant:
$$ W_\mu  = W^{(0)}_\mu + \tilde W_\mu
  = m^{-1/2} \sum_{M= -\hat J}^{\hat J} k_M \psi^M_\mu + \tilde W_\mu  \,.
     \eqn\Wexpand $$
The actual values of the $k_M$ reflect a  balancing of the effects of the
quadratic terms in the action against those of the dominant higher order
terms.  For identifying which terms are dominant, it is useful to define a
dimensionless quantity $a$ by
$$  \sum_{M=-\hat J}^{\hat J} |k_M|^2 =  a^2 \, .
    \eqno\eq $$
Thus, $W^{(0)}_\mu$ is of order $a$.  The source for the
electromagnetic perturbations is quadratic in $W_\mu$ and contains an
explicit factor of $e$, implying that $\delta A_\mu = O(ea^2)$.  The
leading perturbation to the energy-momentum tensor is of order $a$,
arising both from terms quadratic in $W_\mu$ and terms linear in
$\delta A_\mu$.  These enter Einstein's equations multiplied by a
factor of $G$, implying that the metric perturbations are of order
$Gm^2a^2$, where the $m^2$ follows from dimensional considerations.
Substituting these back into the $W_\mu$ field equations, one finds that
$\tilde W$ is of order $e^2a^3$ and thus subdominant, as expected, provided
that $a$ is small.

We can now use these estimates to identify the relevant terms in the
action.  If we assume that $Gm^2$ is small, and in particular that
$Gm^2 \ll a^2$, it is sufficient to consider only those terms that are of
up to order $e^2a^4$.   These involve only $W^{(0)}_\mu$ and
$\delta A_\mu$.  To leading order, the latter is determined in terms of
$W^{(0)}_\mu$ by the
linearized version of the electromagnetic field equation,
$$ {1\over \sqrt{\bar g}} \partial_\mu \left(\sqrt{\bar g}\,
   \delta F^{\mu\nu} \right) =
    {1\over \sqrt{\bar g}} \partial_\mu \left(\sqrt{\bar g}\,
         p^{\mu\nu} \right) +j^\nu
   \eqn\Feq $$
where
$$  p_{\mu\nu} = -{ieg\over 2} \left(W^{(0)*}_\mu W^{(0)}_\nu
     -W^{(0)*}_\nu W^{(0)}_\mu
       \right)
  \eqno\eq $$
and
$$ j^\nu = ie \left[W^{(0)*}_\mu
   \left(\bar D^\mu W^{(0)\nu}u -\bar D^\nu W^{(0)\mu} \right)
      - W^{(0)}_\mu \left(\bar D^\mu W^{(0)\nu *}
   -\bar D^\nu W^{(0)\mu *} \right)  \right] \, .
   \eqno\eq $$
With the aid of these equations, as well as Eqs.~\eigenvalue\ and \psinorm,
the relevant terms in the action may be written as
$$  I =  -\beta^2 m a^2
    +  \int d\,^3x \sqrt{\bar g} \left[
       {\lambda e^2\over 4} \left| W^{(0)*}_\mu W^{(0)}_\nu
     -W^{(0)*}_\nu W^{(0)}_\mu\right|^2
      -{1\over 4} \delta F_{\mu\nu} p^{\mu\nu}
    + {1\over 2}\delta A_\nu  j^\nu \right]\, .
    \eqn\Igeneraleq $$
Since $\delta A_\mu$ should be understood here to be given in terms of
the $k_M$ through Eq.~\Feq, $I$ is simply a polynomial in the $k_M$.  The
leading approximation to $W_\mu$ corresponds to the choice for the $k_M$ that
minimizes this polynomial.   Using the order of magnitude estimates given
above, together with the fact that the typical spatial scale of the
$\psi^M_\mu$ is $m^{-1}$, one finds that the contribution from the first term
in the integral in Eq.~\Igeneraleq\ is of order $\lambda e^2 m a^4$, while
that from the remaining terms in the integral is of order $e^2 m a^4$.
It follows that  the $k_M$ that minimize $I$ will give a value for $a$ that is
of order $\beta/e \sqrt{\lambda +1} $, and thus can be made arbitrarily small
by choosing $r_H$ sufficiently close to $r_{\rm cr}$.

     Having thus determined $W^{(0)}_\mu$ and the leading approximation to
$\delta A_\mu$, one can use the linearized Einstein equations to obtain the
metric perturbations to first order, and then iterate the equations to
obtain the higher order corrections to the various fields.  These results are
described in detail elsewhere\refmark{\construction}.  I will not go into
them any further here, but will instead spend the remainder of this lecture
discussing the symmetry of these solutions.

     We have seen that if $q\ne 1$ there are no $J=0$ vector spherical
harmonics, and hence that the solution cannot be spherically symmetric.  This
still leaves the possibility that the black hole might possess an axial
symmetry.  We can decide whether this is the case by examining  the relative
magnitudes of the various $k_M$.  If the solution is axially symmetric we
can, without any loss of generality, take the $z$-axis to be the axis of
symmetry.  The only possibilities then are either that only $k_0$ is
nonvanishing, in which case the solution is manifestly invariant under
rotations about the $z$-axis, or else that a single one of the other $k_M$ is
nonvanishing, in which case the solution is unchanged if such a rotation is
supplemented by a gauge transformation.   Thus, the new black hole
solution is axially symmetric if and only if
Eq.~\Igeneraleq\ can be minimized with only a single nonzero $k_M$.

     Because the first term on the right hand side depends only on the overall
scale $a$, but not on the relative magnitudes of the various $k_M$, we need
only examine the integral on the right hand side of Eq.~\Igeneraleq.   This
is particularly simple in the case where $\lambda$ is large, so that the
integral is dominated by its first term, and  $g\ge 0$
and $q \ne 1/2$, so that the unstable mode has $J=q-1$.  Because
there is only a single vector harmonic with this total
angular momentum, $W^{(0)\mu}$ then
involves only a single radial function and is of the form
$$   W^{(0)\mu} = f(r)  \sum_{M= -\hat J}^{\hat J} k_M
  C_{q,q-1,M}^\mu \, .
  \eqno\eq $$
The integral in Eq.~\Igeneraleq\ can then approximated by
$$   I' = {\lambda e^2 \over 2}  P(k_M)
    \int_{r_H}^\infty dr \,{f(r)^4 \over r^2} \,
    \eqno\eq $$
where $P(k_M)$ is a quartic polynomial whose coefficients are
obtained from the angular integration of products of
four vector harmonics.  The problem of determining the symmetry of the
solution is thus reduced to the minimization of a polynomial in $2J +1 =
2q-1$ complex variables, subject to the constraint that the sums of their
squares be held fixed.  For $q=2$, one finds that the minimum can be
achieved with a  single nonzero $k_M$, showing that the solution is indeed
axially symmetric.
However, when $q>2$ this axial symmetry disappears; the static
solutions with these higher charges  have no continuous rotational
symmetry

Thus, to summarize, we have found that theories with charged vector
mesons admit a rich variety of magnetically charged black holes that
provide counterexamples to generalizations suggested by the properties
of black holes in simpler theories.  These new solutions include black
holes with hair, new types of extremal black holes with repulsive
mutual forces, and, finally, the first examples of static black hole
solutions that have no rotational symmetry.

\vskip1truecm \noindent{\fourteenbf Acknowlegments}\vskip0.5truecm

I would like express my appreciation to the organizers for arranging
a very enjoyable conference in a beautiful setting.  I would also like
to thank my collaborators, Kimyeong Lee, Parameswaran Nair, and Alex
Ridgway.

\def\refout{
   \ifreferenceopen \Closeout\referencewrite \referenceopenfalse \fi
   \line{\fourteenbf\noindent References\hfil}\vskip\headskip
   \input \jobname.refs   }
\vskip1truecm
\refout
\end